\documentclass[12pt]{article}           
\usepackage[margin=1in]{geometry}                               
\usepackage{graphicx}                           

\usepackage{amssymb}
\usepackage{amsmath}
\usepackage{amsthm}
\usepackage{dsfont}
\usepackage{textcomp}
\usepackage{multirow}

\usepackage{breqn}
\usepackage{bm}
\usepackage{natbib}
\usepackage{color}
\usepackage{setspace}
\usepackage{authblk}
\usepackage{hyperref}
\usepackage{soul}

\usepackage[framemethod=tikz]{mdframed}

\newtheorem{theorem}{Theorem}

\newtheorem{proposition}{Proposition}

\hypersetup{
   bookmarks=true,         
   unicode=false,          
   pdftoolbar=true,        
   pdfmenubar=true,        
   pdffitwindow=false,     
   pdfstartview={FitH},    
   pdftitle={My title},    
   pdfauthor={Author},     
   pdfsubject={Subject},   
   pdfcreator={Creator},   
   pdfproducer={Producer}, 
   pdfkeywords={keyword1} {key2} {key3}, 
   pdfnewwindow=true,      
   colorlinks=true,       
   linkcolor=red,          
   citecolor=blue,        
   filecolor=magenta,      
   urlcolor=blue          
}

\newcommand{\mjfan}[1]{\textcolor{black}{#1}}

\title{\bf{Modeling Tangential Vector Fields on a Sphere}}
\author[1]{Minjie Fan \thanks{Corresponding author.}}
\author[1]{Debashis Paul}
\author[1]{Thomas C.M. Lee}
\affil[1]{Department of Statistics, University of California, Davis}
\author[2]{Tomoko Matsuo}
\affil[2]{Cooperative Institute for Research in Environmental Sciences, University of Colorado, Boulder}
\date{}                                                  

\begin{document}
\maketitle

\begin{abstract}
Physical processes that manifest as tangential vector fields on a sphere are common in geophysical and
environmental sciences. These naturally occurring vector fields are often subject to
physical constraints, such as being curl-free or divergence-free.
We construct a new class of parametric models for cross-covariance functions of
curl-free and divergence-free vector fields that are tangential to the unit sphere.
These models are constructed by applying the surface gradient or the surface curl operator
to scalar random potential fields defined on the unit sphere.
We propose a likelihood-based estimation procedure for the model parameters and show that
fast computation is possible even for large data sets when the observations
are on a regular latitude-longitude grid. Characteristics and utility of the proposed methodology are
illustrated through simulation studies and by applying it to an ocean surface wind
velocity data set collected through satellite-based scatterometry remote sensing. We also compare
the performance of the proposed model with a class of bivariate Mat\'{e}rn models in terms of
estimation and prediction, and demonstrate that the proposed model is superior in capturing
certain physical characteristics of the wind fields.
\end{abstract}


\section{Introduction}\label{sec:intro}


Vector fields defined on a spherical domain are principal objects of study in many branches of science.
Terrestrial physical processes such as wind and oceanic currents, gravity, electric and magnetic fields are some of the most
well-studied examples. In meteorology, the directionality of wind flow at surface level is an example
of a tangential vector field defined on a sphere (Earth's surface).
Because the effective portion of the atmosphere and the oceans are considerably thinner in their vertical extent in
comparison with their horizontal extent, for many geophysical processes, the horizontal scale far exceeds the
vertical scale. It is therefore natural to decompose such vector fields into tangential and radial
components and treat them separately. For example, velocity divergence is expressed
as the sum of horizontal (tangential) and vertical (radial) divergence in the
continuity equation. Other such examples of vector fields include electric and magnetic fields in geophysics \citep{Sabaka-10}. 
A branch of solid-Earth geophysics focuses on the behavior of the Earth's magnetic field, while its interactions with the solar wind are the focus of science of space weather. 
Electric and magnetic fields associated with ionospheric electric currents are distinct in the tangential direction from the radial direction, and many of the ionospheric electrodynamic processes can be treated as tangential vector fields on a sphere \citep{Richmond-88}.
In this paper, we focus on modeling the tangential
component of vector fields on the unit sphere, especially its small-scale stochastic
variations, which can form a significant part of the overall variability.
Accounting for this variability can enhance prediction accuracy of models,
enable subgrid-scale inference, and facilitate better uncertainty quantification of model parameters.
The modeling framework that we propose is flexible, incorporates small-scale variations,
and is naturally interpretable and computationally tractable.

Gaussian random fields (GRF) have provided a very successful modeling framework for describing
variations in many physical processes. Two key ingredients of the success of GRF modeling for
stochastic processes are: (i) the behavior of the process is entirely characterized 
by the mean and the covariance functions, thereby facilitating deep theoretical investigations; and
(ii) computations for both estimation and prediction primarily involve matrix algebra.
We provide a very brief overview of existing literature on the construction of covariance functions
and random fields on a sphere.
\cite{MarinucciP2011} gave a detailed account of 
weakly isotropic processes on the unit sphere, while
\cite{GuinnesF2015} gave a broad overview of spherical processes.
\cite{HitczenkoS2012} \cite{Jun-11}, \cite{JunS2007}, \cite{JunS2008} mostly focused on the construction and characterization of covariance functions of Gaussian processes on the unit sphere.

Extension of the modeling framework from scalar random fields to random vector fields poses an additional challenge due to the requirement of non-negative definiteness for the cross-covariance function
of the vector fields.
Popular approaches to modeling vector fields include linear coregionalization models \citep{Bourgault-91,GelfandSBS2004,Goulard-92},
the multivariate Mat\'ern model \citep{Apanasovich-12, Gneiting-10}, and models derived from scalar
potentials through differential operators \citep{Jun-11,Jun-14}; the latter being
especially notable for their constructive approach.
However, these models do not impose specific physical constraints on the Cartesian components of the vector fields. There are many examples of vector fields satisfying ``natural'' physical constraints.
For example, magnetic fields are solenoidal with zero divergence, and electric fields resulting from electric charges are curl-free.
Vorticity, defined as the curl of a velocity vector field, plays an important role in atmospheric and
oceanic dynamics in terms of characterizing the nature and degree of turbulence. \mjfan{Assuming water is incompressible, \citet{Zhang-07} represented lake water current velocity fields through applying the curl operator to a scalar stream function. \citet{Constantinescu-13} developed cross-covariance models that incorporate known physical constraints relating the behavior of their Cartesian components, 
and demonstrated that physics-based models can significantly outperform independence models in an application to geostrophic winds.
These examples indicate that kriging or data assimilation for observations of geophysical and environmental processes will benefit from modeling vector fields as Gaussian processes that incorporate constraints such as curl-free or divergence-free into their construction.}

\begin{figure}[htbp] 
   \centering
   \includegraphics[width=5.5in]{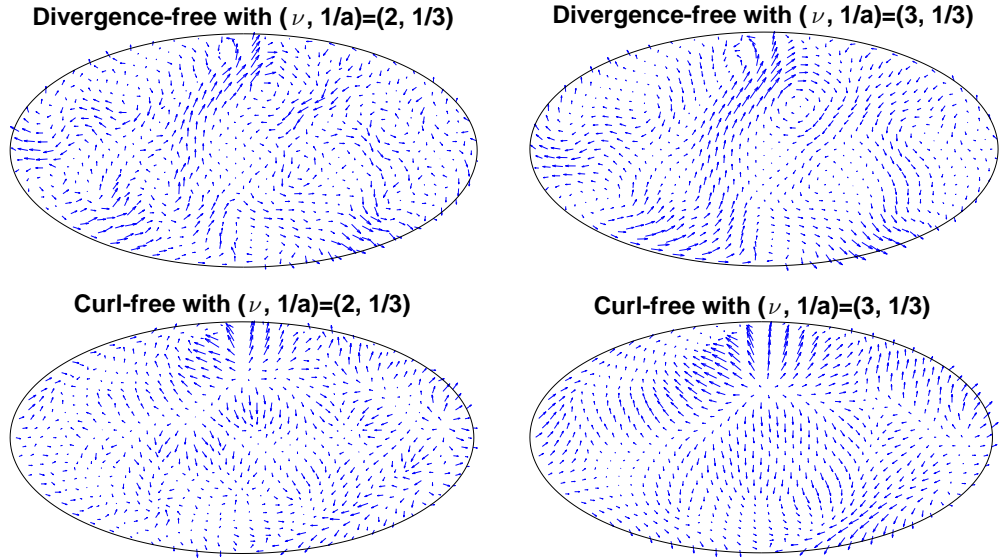}
   \caption{Simulated divergence-free and curl-free random fields on the HEALPix grid. The sphere has been projected to an ellipse by the Hammer projection. Using the same set of random deviates, these fields are simulated based on the Cholesky decomposition. \mjfan{The details will be described in Section \ref{simulation}}.}
   \label{fig:sim}
\end{figure}

\mjfan{We illustrate the basic characteristics of a vector field with physical constraints through Figure \ref{fig:sim}, which shows an example of a pair of simulated divergence-free and curl-free random fields
using models that will be presented later in this paper. The details of how these fields are simulated will be described in Section \ref{simulation}.}

Modeling and analysis of
vector fields (not necessarily tangential to a sphere)
is facilitated by the celebrated Helmholtz-Hodge decomposition \citep{Freeden-09,Harouna-12}
that expresses the vector field as the sum of a curl-free component, a divergence-free component
and a component that is the gradient of a harmonic function. The
harmonic component vanishes when the domain is a spherical surface and the vector field is tangential to this surface. In Section \ref{div_curl_model}, we define the basic objects such as surface gradient and surface curl operators that are fundamental to the construction of a curl-free or a divergence-free component.
Our work is related to that of \citet{Scheuerer-12}, who constructed random
vector fields that are either curl-free or divergence-free, and  that of
\citet{Schlather-15} who introduced a parametric cross-covariance model for random vector fields consisting of the aforementioned curl-free and divergence-free components.
However, these constructions are restricted to vector fields defined on Euclidean spaces only.

In this paper, we construct Gaussian tangential vector fields on the unit sphere that are either curl-free or divergence-free. We then utilize the Helmholtz-Hodge decomposition
to construct general Gaussian tangential vector fields from a pair of correlated Gaussian scalar potential
fields. These constructions involve making careful use of the spherical geometry
and appropriate differential operators on the unit sphere.
\mjfan{Through specifying a bivariate Mat\'ern model for the pair of potential fields,} we propose a simple but flexible parametric model named Tangent Mat\'{e}rn Model (TMM)
with the following characteristics:
\begin{itemize}
\item[(i)] There is a parameter controlling the correlation between the curl-free
and the divergence-free components of the model, which is often non-zero for wind velocity fields \citep{Cornford-98}.
\item[(ii)] The vector field can be equivalently represented in terms of its zonal and meridional components,
and in this form, it allows for a negative correlation between two locations, which is common in meteorological variables
\citep[Chapter 4.3]{Daley-91}.
\item[(iii)] \mjfan{The vector field is modeled as the sum of a curl-free and a divergence-free components. The 
magnitude of each component, which is related to two important physical characteristics of the vector field, divergence and vorticity, can be inferred from the parameter estimates.}
\item[(iv)] The cross-covariance structure of the vector field is \mjfan{not isotropic but} axially symmetric.
\end{itemize}
We then propose a likelihood-based estimation procedure for the TMM and study its characteristics
in terms of estimation through a simulation study. We also present a fast computational
algorithm that is applicable when the data are observed on a regular latitude-longitude grid.
The proposed modeling framework, which is primarily used to describe the small-scale features of a vector field,
is also extended to describe large-scale and spatio-temporal variability.

As an illustration, we apply the proposed methodology to a data set on ocean surface wind velocities
obtained from a satellite survey conducted by NASA's Quick Scatterometer (QuikSCAT) satellite.
Scatterometer data are important for numerical weather prediction. Through the data assimilation
process of surface winds, the analysis of the atmospheric mass and motion fields above the surface can be improved, which increases the accuracy of weather forecasts. Scatterometer data also contribute to improved storm warning and monitoring \citep{Atlas-01, Brennan-09}.
Over the past two decades, a number of statistical methods have been applied to modeling wind fields. For example, \citet{Wikle-99}
demonstrated that enhancement in prediction accuracy of spatio-temporal Kalman filtering
is possible through careful incorporation of small-scale stochastic variations into the model. \citet{Wikle-01} proposed a spatio-temporal hierarchical Bayesian model for tropical ocean surface winds, where the model includes small-scale wind components that are represented in terms of wavelet basis functions.
\citet{Reich-07} developed a multivariate semiparametric Bayesian spatial model for
hurricane surface wind fields based on the stick-breaking prior, while \citet{Foley-08} applied a linear coregionalization model to the zonal and meridional components of hurricane surface wind fields.
Also, \citet{Cornford-04} fitted a version of the model of \citet{Schlather-15} using a Mat\'{e}rn covariance function
to surface wind fields over a small sector of the north Atlantic ocean, by making planar
approximation to the spherical surface.
Due to the many-faceted complexity of the data, intricacies of the available scientific models, and space limitation,
full-scale modeling of surface wind velocity fields is beyond the scope of this paper. Instead, we focus here
primarily on the characteristics of the residual fields after removing large-scale components
through an empirical orthogonal function (EOF) analysis, and demonstrate the implications of
using the TMM for fitting such fields.  We also compare the predictive performance
of co-kriging based on the proposed TMM and a parsimonious bivariate Mat\'ern model
\citep{Gneiting-10}. The results show that the TMM performs better in terms of both estimation and prediction,
which can be attributed to its ability of capturing the physical characteristics of the wind fields.


The remainder of the paper is organized as follows. In Section \ref{div_curl_model}, we present
the constructions of curl-free and divergence-free random tangential vector
fields and the TMM. The details of model fitting are also discussed in this section.
\mjfan{We conduct numerical experiments in Section \ref{sim} to show the good performance of parameter
estimation and spatial prediction.} The TMM is applied to a QuikSCAT ocean surface wind velocity data set  in Section
\ref{real}. We discuss some relevant issues and future directions in Section \ref{dis}.

\section{Construction of Random Tangential Vector Fields}\label{div_curl_model}

\subsection{Surface Gradient and Surface Curl Operators}

Tangential vector fields on the unit sphere can be constructed by applying
appropriate differential operators to scalar potential fields.
Let $\mathbf{s}$ denote a point on the unit sphere $\mathbb{S}^2=\{ \mathbf{x}\in \mathbb{R}^3:
\lVert \mathbf{x} \rVert=1\}$ and $(\theta,\phi)$ represent
the same in the spherical coordinate system, where $\theta$ and $\phi$ are the co-latitude and longitude, respectively.
The tangent space at $\mathbf{s} \equiv (\theta, \phi)$, to be denoted by ${\cal T}_{\mathbf{s}}$,
is a two-dimensional vector space with canonical orthonormal basis vectors $\hat{\bm{\theta}}$ and $\hat{\bm{\phi}}$.
These two vectors are defined such that $\hat{\bm{\theta}}$ is tangent to the great circle
passing through the poles and $\mathbf{s}$, and it points to the south pole, while
$\hat{\bm{\phi}}$ points eastward along the circle that is the intersection of $\mathbb{S}^2$ with a plane passing through $\mathbf{s}$ and parallel to the equatorial plane.
Moreover, let $\hat{\mathbf{r}}$ be a unit vector at $\mathbf{s}$ that points radially outward,
i.e., $\hat{\mathbf{r}}=\hat{\bm{\theta}} \times \hat{\bm{\phi}}$, where $\times$ denotes
the cross product on $\mathbb{R}^3$. All the vectors vary with $\mathbf{s}$, but for the
moment we suppress this.
In order to define spherical differential operators based on
usual differential operators on $\mathbb{R}^3$, for any function or random field on
$\mathbb{S}^2$ mentioned hereafter, we always assume that it is actually defined on
some spherical shell containing $\mathbb{S}^2$,
$S_{\epsilon}=\{ \mathbf{x}\in \mathbb{R}^3: 1-\epsilon<\lVert \mathbf{x} \rVert < 1+\epsilon\}$,
where $\epsilon>0$.

For the unit sphere, the surface gradient of a continuously differentiable function
$f :\mathbb{S}^2 \to \mathbb{R}$ is defined as the tangential component of its usual gradient
\begin{equation*}\label{eqn_P}
\nabla_{\mathbf{s}}^{*}f=:\mathbf{P}_{\mathbf{s}} \nabla_{\mathbf{s}} f,
\end{equation*}
where $\mathbf{s}=(s_1, s_2, s_3)^{\rm T}\in \mathbb{S}^2, \mathbf{P}_{\mathbf{s}}=\mathbf{I}_3-\mathbf{s}\mathbf{s}^{\rm T}$
is the matrix that projects a vector in $\mathbb{R}^3$ onto $\mathcal{T}_{\mathbf{s}}$,
and $\nabla_{\mathbf{s}}$ is the usual gradient. Then the surface curl of $f$ is defined as
a tangential vector field perpendicular to $\nabla_{\mathbf{s}}^{*}f$
\begin{equation*}\label{eqn_Q}
L_{\mathbf{s}}^{*}f=:\hat{\mathbf{r}} \times \nabla_{\mathbf{s}}^{*}f=\mathbf{Q}_{\mathbf{s}} \nabla_{\mathbf{s}} f,
\end{equation*}
where the matrix $\mathbf{Q}_{\mathbf{s}}$ is defined in (\ref{eq:Q_s}) of Appendix A.

If we further assume that $f$ is twice continuously differentiable,
then  $\nabla^*_{\mathbf{s}}f$ and  $L_{\mathbf{s}}^*f$ are curl-free and divergence-free, respectively, i.e.,
$\mathrm{curl}_{\mathbf{s}}^*(\nabla^*_{\mathbf{s}}f)=: L_{\mathbf{s}}^* \cdot \nabla^*_{\mathbf{s}}f=0$,
where $\cdot$ denotes the dot (or scalar) product,
and $\mathrm{div}_{\mathbf{s}}^*(L_{\mathbf{s}}^*f )=:\nabla_{\mathbf{s}}^* \cdot L_{\mathbf{s}}^*f=0$.
If $f$ is only continuously differentiable, $\mathrm{curl}_{\mathbf{s}}^*(\nabla^*_{\mathbf{s}}f)$
and $\mathrm{div}_{\mathbf{s}}^*(L_{\mathbf{s}}^*f )$ are not well-defined in the ordinary sense,
and hence we call $\nabla^*_{\mathbf{s}}f$ the \textit{gradient-field} and $L_{\mathbf{s}}^*f$
the \textit{curl-field} derived from $f$. More details on differential calculus on the unit sphere can
be found in Chapter 2 
of \citet{Freeden-09}.
\mjfan{Some auxiliary notations and definitions are given in Appendix A.}

\subsection{Construction of Curl-free and Divergence-free Random Tangential Vector Fields}\label{construct}

In this subsection, we present an approach for constructing
curl-free and divergence-free random tangential vector fields on the unit sphere. These vector fields
are derived by applying the surface gradient and the surface curl operators to a scalar random potential field, to be generically denoted by $Z(\mathbf{s})$. The cross-covariance structure of
the derived vector fields will be determined by the covariance structure of the process $Z$.
Suppose $Z$ is defined on $S_\epsilon$ with mean zero and finite variance, and
satisfies the following regularity conditions:
\begin{enumerate}
\item[\textbf{A1}] The process $Z(\mathbf{s})$ is differentiable in quadratic mean. Moreover, there
exists a $\mathbb{P}-$a.e. sample continuously differentiable version of $Z$, denoted by $\widetilde{Z}$, such that $D^{(i)}\widetilde{Z}(\mathbf{s})=D_{\mathrm{qm}}^{(i)}Z(\mathbf{s})$ $\mathbb{P-}$a.e.\mjfan{, where $D^{(i)}$ and $D^{(i)}_{\rm qm}$ represent the sample partial derivative and the partial derivative in quadratic mean along the $i$-th coordinate direction, respectively.}
\item[\textbf{A2}] The process $Z(\mathbf{s})$ is stationary with twice continuously differentiable
covariance function $C(\mathbf{h})$, where $\mathbf{h}$ is the spatial separation vector
between $\mathbf{s}$ and $\mathbf{t}$, i.e., $\mathbf{h}=\mathbf{s}-\mathbf{t}$.
\end{enumerate}
The two regularity conditions ensure the validity of applying the differential operators.
When $Z(\mathbf{s})$ is Gaussian, \textbf{A1} can be verified through Theorems 3.2 and 4.3 in \citet{Potthoff-10}.
By \textbf{A1}, there exists $\Omega_0 \subset \Omega$ with $\mathbb{P}(\Omega_0)=1$
such that $\widetilde{Z}(\mathbf{s}, \omega)$ is continuously differentiable on $S_{\epsilon}$
for any $\omega \in \Omega_0$. By applying the differential operators to $\widetilde{Z}$, which can be seen as a scalar random potential field,
we construct two random tangential vector fields $\mathbf{Y}_{\mathrm{curl}, Z}$ and
$\mathbf{Y}_{\mathrm{div}, Z}$ on $\mathbb{S}^2$ such that  $\mathbf{Y}_{\mathrm{curl}, Z}(\mathbf{s}, \omega)
=\mathbf{P}_{\mathbf{s}}\nabla_{\mathbf{s}}\widetilde{Z}(\mathbf{s}, \omega)$ and $\mathbf{Y}_{\mathrm{div}, Z}(\mathbf{s}, \omega)=\mathbf{Q}_{\mathbf{s}}\nabla_{\mathbf{s}}\widetilde{Z}(\mathbf{s}, \omega)$ for any $\omega \in \Omega_0$.
It is clear that the sample paths of $\mathbf{Y}_{\mathrm{curl}, Z}$ and $\mathbf{Y}_{\mathrm{div}, Z}$ are gradient-fields and curl-fields $\mathbb{P}-$a.e., respectively.
If we further assume that the sample paths of $\widetilde{Z}$ are twice continuously differentiable $\mathbb{P}-$a.e.,
then $\mbox{curl}_{\mathbf{s}}^*(\mathbf{Y}_{\mathrm{curl}, Z}(\mathbf{s}, \omega))$ and $\mbox{div}_{\mathbf{s}}^*(\mathbf{Y}_{\mathrm{div}, Z}(\mathbf{s}, \omega))$ are well-defined and equal to zero for any $\omega \in \Omega_0$,
and hence $\mathbf{Y}_{\mathrm{curl}, Z}$ and $\mathbf{Y}_{\mathrm{div}, Z}$ are curl-free and divergence-free, respectively.

The cross-covariance functions of  $\mathbf{Y}_{\mathrm{curl}, Z}$ and $\mathbf{Y}_{\mathrm{div}, Z}$,
to be denoted by $\mathbf{C}_{\mathrm{curl}, Z}$ and $\mathbf{C}_{\mathrm{div}, Z}$, respectively,
are given explicitly in Theorem \ref{main_thm}, the proof of which appears in Appendix B.
\begin{theorem}\label{main_thm}
If \textbf{A1} and \textbf{A2} hold, then the cross-covariance functions
$\mathbf{C}_{\mathrm{curl}, Z}$ and $\mathbf{C}_{\mathrm{div}, Z}$ can be represented as
\begin{equation}\label{curl_cov}
\mathbf{C}_{\mathrm{curl}, Z}(\mathbf{s}, \mathbf{t})=-\mathbf{P}_{\mathbf{s}} \nabla_{\mathbf{h}} \nabla_{\mathbf{h}}^{\rm T}C(\mathbf{h}) \bigg|_{\mathbf{h}=\mathbf{s}-\mathbf{t}} \mathbf{P}_{\mathbf{t}}^{\rm T},
\end{equation}
and
\begin{equation}\label{div_cov}
\mathbf{C}_{\mathrm{div}, Z}(\mathbf{s}, \mathbf{t})=-\mathbf{Q}_{\mathbf{s}} \nabla_{\mathbf{h}} \nabla_{\mathbf{h}}^{\rm T}C(\mathbf{h}) \bigg|_{\mathbf{h}=\mathbf{s}-\mathbf{t}} \mathbf{Q}_{\mathbf{t}}^{\rm T}.
\end{equation}
\end{theorem}


We consider a special case of the construction described above, where $Z(\mathbf{s})$ is an \textit{isotropic} scalar random field on $S_{\epsilon}$,
so that its covariance function $C(\mathbf{h})$ can be written as $C_1(\lVert \mathbf{h} \rVert)$, where $C_1(r)$ is a function from
$[0, 2+2\epsilon)$ to $\mathbb{R}$. One popular choice for $C_1$ is the Mat\'ern model
\begin{equation}\label{matern}
M(r;\nu, a)=\frac{2^{1-\nu}}{\Gamma(\nu)}(ar)^\nu K_{\nu}(ar),
\end{equation}
where $K_{\nu}$ is the modified Bessel function of the second kind, the parameter $\nu>0$
controls the smoothness, and $a>0$ is the spatial scale parameter, whereby $1/a$ controls the range  of correlation. When $Z(\mathbf{s})$
is assumed to be Gaussian with covariance function $C_1(\lVert \mathbf{h} \rVert)=M(\lVert \mathbf{h} \rVert ;\nu, a)$,
$\nu>1$ is necessary and sufficient to ensure that conditions \textbf{A1} and
\textbf{A2} hold. We give a proof of the statement in Appendix C, and explicit expressions of $\nabla_{\mathbf{h}} \nabla_{\mathbf{h}}^{\rm T}C_1(\lVert \mathbf{h} \rVert)$ in Appendix D.

We now establish connections of our proposal with existing literature.
\citet{Narcowich-07} and \citet{Edward-09} constructed a general class of positive definite
curl-free and divergence-free kernels on $\mathbb{S}^2$ using radial basis functions (RBFs)  on $\mathbb{R}^3$.
The cross-covariance kernels defined in (\ref{curl_cov}) and (\ref{div_cov}), when $Z(\mathbf{s})$ is isotropic,
belong to this class. However, the constructions of \citet{Narcowich-07} and \citet{Edward-09} are
deterministic in nature, and the descriptions of their kernels do not automatically
lead to the construction of random tangential vector fields. In contrast, our construction takes
a different path, by directly deriving vector fields through the application of differential operators to
scalar random potential fields. Moreover, we can extend the basic
framework presented here to construct a rich class of spatio-temporal tangential vector fields on the unit sphere 
with more complicated cross-covariance structures and including large-scale components. These ideas will be developed in {two stages}, first by constructing
a tangential vector field with correlated curl-free and divergence-free components from
a pair of correlated scalar random potential fields (Section \ref{para_model}), and then by
extending the framework to spatio-temporal processes (Section \ref{spa_temp_framework}). 


\subsection{Tangent Mat\'ern Model}\label{para_model}

In this subsection, we construct a class of Gaussian tangential vector fields by making use of the
Helmholtz-Hodge decomposition and the ingredients developed in Section \ref{construct}.
Let $\mathbf{Z}(\mathbf{s})=(Z_1(\mathbf{s}), Z_2(\mathbf{s}))^{\rm T}$ denote an isotropic, bivariate, Gaussian random field
on $S_{\epsilon}$ with mean zero and cross-covariance function $\mathbf{C}(\lVert \mathbf{h} \rVert) = \mbox{Cov}(\mathbf{Z}(\mathbf{s}),\mathbf{Z}(\mathbf{t}))$,
where $\mathbf{h}=\mathbf{s}-\mathbf{t}$ and $\mathbf{s}, \mathbf{t} \in S_{\epsilon}$.
The Helmholtz-Hodge decomposition states that any continuously differentiable tangential vector field
on $\mathbb{S}^2$ can be uniquely decomposed as the sum of a curl-free and a divergence-free
components \citep{Freeden-09}. 
Following this idea, we first apply the spherical differential operators to $Z_1$ and $Z_2$ to
obtain a curl-free  and a divergence-free vector fields, respectively, as described in Section \ref{construct},
and then define a tangential vector field as the sum of these two components. Specifically,
\begin{equation}\label{eq:Y_mix_def}
\mathbf{Y}_{\mathrm{tan}, \mathbf{Z}}(\mathbf{s})= \mathbf{P}_{\mathbf{s}}\nabla_{\mathbf{s}}\widetilde{Z}_1(\mathbf{s})
+ \mathbf{Q}_{\mathbf{s}}\nabla_{\mathbf{s}}\widetilde{Z}_2(\mathbf{s}) \qquad \mathbb{P-}\mbox{a.e.},
\end{equation}
where $\widetilde{Z}_1$ and $\widetilde{Z}_2$ are $\mathbb{P}-$a.e. sample continuously differentiable
versions of $Z_1$ and $Z_2$, respectively. Given that the transformation from
$\mathbf{Z}$ to $\mathbf{Y}_{\mathrm{tan}, \mathbf{Z}}$ is linear, it is deduced that
$\mathbf{Y}_{\mathrm{tan}, \mathbf{Z}}$ is also a Gaussian random field. A highly advantageous
characteristic of this construction is that it allows for a correlation between the
curl-free and the divergence-free components, which is inherited from the underlying bivariate potential field.

\citet{Gneiting-10} introduced a new class of cross-covariance functions called
multivariate Mat\'ern with a flexible correlation structure among its Cartesian components.
We consider a parsimonious bivariate Mat\'{e}rn model for the underlying potential
field $\mathbf{Z}$ due to its simplicity and flexibility. Here, ``parsimonious'' refers to
the fact that both Cartesian components of the bivariate process have the same spatial scale parameter.
\mjfan{\citet{Gneiting-10} has argued that this assumption is not necessarily restrictive based on the fact that 
the parameters $\sigma^2$ and $a$ of a Mat\'ern covariance function (with a fixed smoothness parameter $\nu$)
in dimension $d\leq 3$ cannot be consistently estimated under infill asymptotics \citep{Zhang-04}.} We refer to the resulting tangential vector field model as \textit{Tangent Mat\'{e}rn
Model (TMM)}.
\mjfan{Write the cross-covariance function $\mathbf{C}(\lVert \mathbf{h}\rVert)$ as $\left(C_{ij}(\lVert \mathbf{h} \rVert)\right)_{1\leq i, j\leq 2}$.}
The parsimonious bivariate Mat\'{e}rn model  specifies
\begin{equation*}\label{pars_m1}
C_{ii}(\lVert \mathbf{h} \rVert)=\sigma_i^2M(\lVert \mathbf{h} \rVert;\nu_i, a) \qquad \mbox{for } i=1,2,
\end{equation*}
and
\begin{equation*}\label{pars_m2}
C_{12}(\lVert \mathbf{h} \rVert)=C_{21}(\lVert \mathbf{h} \rVert)=\rho_{12} \sigma_1 \sigma_2 M(\lVert \mathbf{h} \rVert ;(\nu_1+\nu_2)/2, a),
\end{equation*}
where $\sigma_i^2$ are variance parameters,
and $a$ is the spatial scale parameter shared by the Cartesian components $Z_1$ and $Z_2$. The parameter $\rho_{12}$,
which represents a co-located correlation coefficient, controls the correlation between
$Z_1$ and $Z_2$, and through this, also determines the correlation between the curl-free
and the divergence-free components of the vector field $\mathbf{Y}_{\mathrm{tan}, \mathbf{Z}}$.
A necessary and sufficient condition for non-negative definiteness of the cross-covariance
function $\mathbf{C}(\lVert \mathbf{h}\rVert)$  is that
\begin{equation*}
|\rho_{12}|\leq \frac{\Gamma(\nu_1+\frac{3}{2})^{1/2}}{\Gamma(\nu_1)^{1/2}}
\frac{\Gamma(\nu_2+\frac{3}{2})^{1/2}}{\Gamma(\nu_2)^{1/2}}
\frac{\Gamma(\frac{1}{2}(\nu_1+\nu_2))}{\Gamma(\frac{1}{2}(\nu_1+\nu_2)+\frac{3}{2})}.
\end{equation*}
To ensure a sufficient degree of smoothness that enables the application of differential operators,
the smoothness parameters $\nu_1$ and $\nu_2$ are required to be larger than 1.
The derived cross-covariance function of $\mathbf{Y}_{\mathrm{tan}, \mathbf{Z}}(\mathbf{s})$ is
\begin{eqnarray}\label{mixed}
&& \mathbf{C}_{\mathrm{tan}, \mathbf{Z}}(\mathbf{s}, \mathbf{t}) \nonumber\\
&=& -\left( \begin{matrix}
\sigma_1 \mathbf{P}_{\mathbf{s}} & \sigma_2 \mathbf{Q}_{\mathbf{s}}
\end{matrix} \right)\left( \begin{matrix}
\mathbf{K}(\mathbf{h}; \nu_1, a) & \rho_{12}\mathbf{K}(\mathbf{h};(\nu_1+\nu_2)/2, a)\\
\rho_{12}\mathbf{K}(\mathbf{h};(\nu_1+\nu_2)/2, a) & \mathbf{K}(\mathbf{h}; \nu_2, a)
\end{matrix} \right)
\left( \begin{matrix}
\sigma_1 \mathbf{P}_{\mathbf{t}}^{\rm T} \\
 \sigma_2 \mathbf{Q}_{\mathbf{t}}^{\rm T}
\end{matrix} \right),
\end{eqnarray}
where $\mathbf{K}(\mathbf{h};\nu, a)$ denotes the RHS of (\ref{mat_K}) with $F$ and $G$ given by (\ref{fun_F}) and (\ref{fun_G}), respectively, and (\ref{mat_K}), (\ref{fun_F}) and (\ref{fun_G}) are in Appendix D.

\mjfan{We can relax the assumptions on the underlying potential field $\textbf{Z}$ by assuming a full bivariate Mat\'ern model \citep{Gneiting-10} with distinct spatial scale parameters and a more flexible cross-covariance smoothness parameter (see Supplementary Materials S2). A more complicated constraint of the parameters is required to ensure the validity of the model, and thus the corresponding computation would be more expensive.}

\mjfan{In the rest of this paper, for both the methodology and application, we focus on the TMM
described above. Two characteristics of the model will be presented later, which can be easily extended to tangential vector fields derived 
from more general potential fields than the parsimonious bivariate Mat\'{e}rn model.}


So far, the tangential vector fields are represented in terms of Cartesian coordinates.
For convenience, we also give a representation in terms of canonical coordinates in the tangent space.
A tangential vector field $\mathbf{Y}(\mathbf{s})$ can be represented in the canonical coordinates $(\hat{\mathbf{u}}, \hat{\mathbf{v}})$ of the tangent space ${\cal T}_{\mathbf{s}}$
as $\mathbf{V}(\mathbf{s})\equiv (u(\mathbf{s}), v(\mathbf{s}))^{\rm T}$, where $\hat{\mathbf{u}}=\hat{\bm{\phi}}$, $\hat{\mathbf{v}}=-\hat{\bm{\theta}}$, and $u$ and $v$ are the
zonal (eastward) and meridional (northward) components, respectively.
In terms of the canonical coordinates $(\hat{\bm{\theta}},\hat{\bm{\phi}})$
of the tangent space,
$u(\mathbf{s}) = \mathbf{Y}(\mathbf{s}) \cdot \hat{\bm{\phi}}$ and
$v(\mathbf{s}) = - \mathbf{Y}(\mathbf{s}) \cdot \hat{\bm{\theta}}$, where $\cdot$ denotes the dot (or scalar) product.
The cross-covariance function of $\mathbf{V}$ can be obtained by applying a suitable
transformation between the coordinates, which is given in Proposition \ref{prop1}
below. The proof of the proposition appears in Appendix E.

\begin{proposition}\label{prop1}
Let $\mathbf{Y}_{\mathrm{tan}, \mathbf{Z}}(\mathbf{s})$ be the derived vector field in the TMM defined through
(\ref{eq:Y_mix_def}), where $\nu_1,\nu_2 > 1$, and let $\mathbf{V}(\mathbf{s})\equiv (u(\mathbf{s}), v(\mathbf{s}))^{\rm T}$ be its
representation in the canonical coordinates $(\hat{\mathbf{u}}, \hat{\mathbf{v}})$. The cross-covariance function of $\mathbf{V}$
has the following expression
\begin{equation}\label{uv_cov}
\mathbf{C}_{\mathbf{V}}(\mathbf{s}, \mathbf{t})=\mathbf{T}_{\mathbf{s}}\mathbf{C}_{\rm{tan}, \mathbf{Z}}(\mathbf{s}, \mathbf{t})\mathbf{T}_{\mathbf{t}}^{\rm T},
\end{equation}
\mjfan{where the expression of the transformation matrix $\mathbf{T}_{\mathbf{s}}$ is given in the proof.}
When $\mathbf{s}=\mathbf{t}$,
\begin{equation}\label{collocated}
\mathbf{C}_{\mathbf{V}}(\mathbf{s}, \mathbf{s})=-\left[\sigma_1^2F_{\rm{Mat}}(0; \nu_1, a)+\sigma_2^2F_{\rm{Mat}}(0; \nu_2, a)\right]\mathbf{I}_2.
\end{equation}
\mjfan{The co-located correlation between $u$ and $v$ is exactly zero, which is due to the isotropy of the underlying potential field.}
\end{proposition}
For a scalar random field $X(\mathbf{s})\equiv X(\theta_{\mathbf{s}}, \phi_{\mathbf{s}})$ defined on $\mathbb{S}^2$, we say that $X$ is \textit{axially symmetric} if
$$
\mbox{Cov}(X(\theta_{\mathbf{s}}, \phi_{\mathbf{s}}), X(\theta_{\mathbf{t}}, \phi_{\mathbf{t}}))
=C(\theta_{\mathbf{s}},\theta_{\mathbf{t}}, \phi_{\mathbf{s}}-\phi_{\mathbf{t}}),
$$
for any $\theta_{\mathbf{s}}, \theta_{\mathbf{t}}, \phi_{\mathbf{s}}, \phi_{\mathbf{t}}$ and
some function $C$ \citep{Jones-63}. Proposition \ref{prop2} below gives the
representation of $\mathbf{V}(\mathbf{s})$ in the spherical coordinate system, the proof of which appears in Appendix F.
\begin{proposition}\label{prop2}
The random field $\mathbf{V}(\mathbf{s})\equiv (u(\mathbf{s}), v(\mathbf{s}))^{\rm T}$ can be represented (except at the two poles) as
\begin{equation}\label{eqn_u}
u(\theta, \phi)=\frac{1}{\sin{\theta}}\frac{\partial Z_1}{\partial \phi}+\frac{\partial Z_2}{\partial \theta} \qquad \mathbb{P-}\mbox{a.e.},
\end{equation}
and
\begin{equation}\label{eqn_v}
v(\theta, \phi)=\frac{1}{\sin{\theta}}\frac{\partial{Z_2}}{\partial \phi}-\frac{\partial Z_1}{\partial \theta} \qquad \mathbb{P-}\mbox{a.e.},
\end{equation}
for any $\theta \in (0, \pi)$.
Here the partial derivatives are defined in the sense of quadratic mean. It implies that $u$ and $v$
are axially symmetric both marginally and jointly.
\end{proposition}

The spherical representation suggests a similarity between the TMM
and the non-stationary covariance and cross-covariance models introduced by \citet{Jun-11, Jun-14, JunS2008}.
Both of them are represented as a linear combination of partial derivatives of scalar random fields,
and both share the property of axial symmetry.
However, there are also some important differences: (i) Jun's model is derived from a different
perspective, which is to capture global non-stationarity with respect to latitude;
(ii) In Jun's model, the coefficients of
the partial derivatives with respect to $\theta$ and $\phi$ are linear combinations of Legendre polynomials, i.e.,
$\sum_{j=0}^m a_j P_j(\sin (\pi/2-\theta))$, where $P_j$ denotes the Legendre polynomial of order $j$. \citet{Jun-11}
assumes that $Z_1$ and $Z_2$ are uncorrelated, while \citet{Jun-14} extends the model to allow for correlated $Z_1$ and $Z_2$ with spatially varying variance and smoothness parameters using the formulation of \citet{Kleiber-12}.

\subsection{Fast Parameter Estimation Using DFT}\label{para_est}

Suppose $\mathbf{Y}(\mathbf{s}) = (Y_1(\mathbf{s}), Y_2(\mathbf{s}))^{\rm T}$ is a bivariate Gaussian random field
with mean zero and cross-covariance function given by (\ref{uv_cov}). The observations on the process $\mathbf{Y}$
at $n$ different locations are expressed as the vector
$\mathbf{Y}=(\mathbf{Y}(\mathbf{s}_1)^{\rm T}, \cdots, \mathbf{Y}(\mathbf{s}_n)^{\rm T})^{\rm T}$, which has a multivariate normal distribution
of dimension $2n$. Henceforth, we use the notation $\mathbf{Y}$ to denote a random field and a random vector
of observations interchangeably. The negative log-likelihood function (ignoring a constant) is
\begin{equation}\label{negloglik}
l(\bm{\theta})=\frac{1}{2} \mbox{log}|\bm{\Sigma}(\bm{\theta})|+\frac{1}{2}\mathbf{Y}^{\rm T} \bm{\Sigma}(\bm{\theta})^{-1}\mathbf{Y},
\end{equation}
where $\bm{\Sigma}(\bm{\theta})$ is the $2n\times 2n$ cross-covariance matrix and $\bm{\theta}$ is the parameter vector.
Maximum likelihood is very commonly used to estimate the parameters in spatial statistical models. 
However, the computation of the MLEs can be very difficult for large data sets since the evaluation of
the likelihood requires $\mathcal{O}(n^3)$ operations. Nonetheless, in our case, if the observations are on a regular latitude-longitude grid,
the discrete Fourier transform (DFT) can be used to speed up the computation.
Regularly spaced observations are common for remote sensing satellite data, numerical weather
model outputs and meteorological reanalyses. \citet{Jun-11} pointed out that as long as the
regular grid covers the full longitude range, and the cross-covariance function is axially symmetric,
the cross-covariance matrix can be transformed using the DFT to a $2\times 2$ block matrix with each
block being a block diagonal matrix. \mjfan{The implementation details are given in Supplementary Materials S1.}
The DFT helps reduce the time complexity of evaluating $l(\bm{\theta})$ to
$\mathcal{O}\left(n(\log{n_{\rm lon}}+n_{\rm lat}^2)\right)$, where $n_{\rm lat}$ and $n_{\rm lon}$
denote the number of latitude and longitude grid points, respectively, and $n_{\rm lat}n_{\rm lon}=n$. When
$\mathcal{O}(n_{\rm lat})=\mathcal{O}(n_{\rm lon})=\mathcal{O}(\sqrt{n})$, the time complexity is
simplified as $\mathcal{O}(n^2)$. A numerical comparison between the methods with and without using
the DFT will be given in Section \ref{accu}. Note that the DFT is not suitable for irregularly
spaced data nor data that are incomplete on a regular grid. In these cases, one may opt to use
approximation methods such as covariance tapering \citep{Furrer-06, Kaufman-08, Bevilacqua-15}.


\subsection{Spatio-temporal Modeling}\label{spa_temp_framework}

The aforementioned idea of constructing random vector fields based on differential operators 
can be extended to construct spatio-temporal models for tangential vector fields on the unit sphere.
The idea is to start with a pair of correlated scalar potential fields on the unit sphere that vary 
in time. Accordingly, let $(\Phi, \Psi)$ be two scalar random processes on $\mathbb{S}^2 \times \mathbb{Z}$,
with the following decompositions
\begin{equation}\label{eq:spatio_temporal_potential}
\Phi(\mathbf{s}, t)=\mu_{\Phi}(\mathbf{s})+\Phi_{L}(\mathbf{s}, t)+\Phi_S(\mathbf{s}, t); \qquad 
\Psi(\mathbf{s}, t)=\mu_{\Psi}(\mathbf{s})+\Psi_{L}(\mathbf{s}, t)+\Psi_S(\mathbf{s}, t),
\end{equation}
where $\mu_{\Phi}$ and $\mu_{\Psi}$ are nonrandom spatial means, $(\Phi_{L}, \Psi_{L})$ and $(\Phi_{S}, \Psi_{S})$ are 
large-scale and small-scale spatio-temporal components, respectively. It is often assumed that large-scale components are independent of small-scale components.  
The large-scale components are modeled by finite linear combinations of nonrandom spatially varying basis functions,
with random coefficients that are functions of time, or through certain parametric random processes
of $(\mathbf{s},t)$. The former is given by
\begin{equation}\label{eq:large_scale_potential}
\Phi_{L}(\mathbf{s}, t)=\sum_{j=1}^{J} a_j(t)\phi_j(\mathbf{s}); \qquad \Psi_{L}(\mathbf{s}, t)=\sum_{k=1}^{K} b_k(t)\psi_k(\mathbf{s}),
\end{equation}
where $\{\phi_j\}_{j=1}^{J}$ and $\{\psi_k\}_{k=1}^{K}$ are nonrandom basis functions, 
and $\{a_j(t)\}_{j=1}^{J}$ and $\{b_k(t)\}_{k=1}^{K}$ are zero-mean time series, typically assumed
to be stationary.
Such models have been used in modeling spatio-temporal processes defined on both Euclidean and spherical domains. 
Possible choices for the basis functions are spherical harmonics \citep{Stein-07}, 
wavelets \citep{Matsuo-11, Nychka-02} and some overcomplete frame of functions \citep{Hsu-12,Nychka-14}.
For the small-scale potential fields, we may assume that the bivariate process
$(\Phi_{S}(\mathbf{s},t),\Psi_{S}(\mathbf{s},t))$ is independent across time, and for any
fixed $t$, it is an isotropic, bivariate, Gaussian random field with mean zero and the same cross-covariance structure.

For simplicity, we may assume that the scalar potential fields are detrended, i.e., $\mu_{\Phi} \equiv 0$ and $\mu_{\Psi} \equiv 0$.
Applying the spherical differential operators to $\Phi$ and $\Psi$, we have
\begin{equation}\label{eq:spatio_temporal_vector_field}
\nabla_{\mathbf{s}}^*\Phi(\mathbf{s}, t)=\sum_{j=1}^{J}a_j(t) \widetilde{\bm{\phi}}_j(\mathbf{s}) + \nabla_{\mathbf{s}}^* \Phi_{S}(\mathbf{s}, t); 
\qquad
L_{\mathbf{s}}^*\Psi(\mathbf{s}, t)=\sum_{k=1}^{K}b_k(t) \widetilde{\bm{\psi}}_k(\mathbf{s}) + L_{\mathbf{s}}^* \Psi_{S}(\mathbf{s}, t),
\end{equation}
where $\widetilde{\bm{\phi}}_j(\mathbf{s})=\nabla_{\mathbf{s}}^*\phi_j(\mathbf{s})$ and 
$\widetilde{\bm{\psi}}_k(\mathbf{s})=L_{\mathbf{s}}^*\psi_k(\mathbf{s})$.
Thus, we obtain the derived vector field
$$
\mathbf{Y}(\mathbf{s}, t)=\sum_{j=1}^{J} a_j(t) \widetilde{\bm{\phi}}_j(\mathbf{s}) 
+ \sum_{k=1}^{K}b_k(t) \widetilde{\bm{\psi}}_k(\mathbf{s})+\nabla_{\mathbf{s}}^* \Phi_{S}(\mathbf{s}, t)
+L_{\mathbf{s}}^* \Psi_{S}(\mathbf{s}, t).
$$
Representing $\mathbf{Y}(\mathbf{s},t)$ in the canonical coordinates $(\hat{\mathbf{u}}, \hat{\mathbf{v}})$ of the tangent space at $\mathbf{s}$, we have
\begin{equation}\label{spa_temp_model}
\mathbf{V}(\mathbf{s})=\mathbf{T}_{\mathbf{s}}\mathbf{Y}(\mathbf{s}, t) =\sum_{j=1}^{J}a_j(t) \left[\mathbf{T}_{\mathbf{s}}\widetilde{\bm{\phi}}_j(\mathbf{s})\right]
+ \sum_{k=1}^{K}b_k(t) \left[\mathbf{T}_{\mathbf{s}}\widetilde{\bm{\psi}}_k(\mathbf{s})\right]+\mathbf{T}_{\mathbf{s}}\left( \nabla_{\mathbf{s}}^* \Phi_{S}(\mathbf{s}, t)+L_{\mathbf{s}}^* \Psi_{S}(\mathbf{s}, t) \right),
\end{equation}
where $\mathbf{T}_{\mathbf{s}}$ is the transformation matrix defined in Proposition \ref{prop1}.
The sum of the first two terms constitutes the large-scale component of the vector field, 
and the last term constitutes the small-scale component. 
For any fixed $t$, the last term is a tangential vector field analogous to the TMM.


\section{Numerical Results}\label{sim}

\subsection{Simulation of Random Fields}\label{simulation}

This subsection presents simulated zero-mean Gaussian random fields with the curl-free and 
the divergence-free cross-covariance functions constructed in Section \ref{construct}. 
The scalar potential field $Z$ follows the Mat\'ern model $M(\lVert \mathbf{h} \rVert; \nu, a)$, and
the sampling locations are on a HEALPix grid \citep{Gorski-05}, which partitions the unit sphere 
into equal area pixels. Compared with a regular grid, the HEALPix grid is more suitable for 
visualization of spherical functions due to the curvature of the sphere. Figure \ref{fig:sim} 
shows simulated divergence-free (in the first row) and curl-free (in the second row) random 
fields on the HEALPix grid with $768$ grid points. The smoothness parameters are 2 and 3 
for the realizations on the left and on the right, respectively. The spatial scale parameter 
$a=3$. As we can see from Figure \ref{fig:sim}, the realizations on the left are rougher than 
those on the right, which is consistent with the parameter specification.

\subsection{Accuracy of Parameter Estimation}\label{accu}

We turn to investigate the accuracy of parameter estimation by a Monte Carlo simulation 
study. The same notation is used here as in Section \ref{para_est}. Specifically, 
we consider a zero-mean bivariate Gaussian random field $\mathbf{Y}(\mathbf{s})$ that follows the TMM (\ref{uv_cov}). Augmented with nugget effects to account 
for observational error, the cross-covariance function hence becomes
\begin{equation}\label{gen}
\mathbf{C}_{\mathbf{Y}}(\mathbf{s}, \mathbf{t})=\mathbf{T}_{\mathbf{s}}\mathbf{C}_{\mathrm{tan}, \mathbf{Z}}(\mathbf{s}, \mathbf{t})\mathbf{T}_{\mathbf{t}}^{\rm T}+
\mathrm{diag} (\tau_1^2 \mathds{1}(\mathbf{s}=\mathbf{t}), \tau_2^2 \mathds{1}(\mathbf{s}=\mathbf{t}) ),
\end{equation}
where the parameter vector \mjfan{$\bm{\theta}=(\sigma_1, \sigma_2, \rho_{12}, \nu_1, \nu_2, 1/a, \tau_1, \tau_2)$}. 
The sampling locations are on a regular grid with latitudes ranging from $-50$ degree to $50$ degree, which follows the same range of the TOMS Level 3 data analyzed in \citet{JunS2008}.

For the evaluation of likelihood functions, we implement the DFT  described in 
Section \ref{para_est} and compare its computational speed with that of the method 
without using the DFT. On a regular laptop with a 2.4 GHz Intel Core i5 processor, 
when the number of latitudes $n_{\rm lat}=25$ and the number of longitudes $n_{\rm lon}=50$, 
the method using the DFT only takes $3.03$ seconds, while the one without using the DFT 
takes $34.43$ seconds. To minimize the negative log-likelihood function (\ref{negloglik}), 
we use the \emph{interior-point} algorithm through the Matlab function \emph{fmincon} due 
to its capability of handling large-scale problems. To specify the initial value of the 
parameters, we first fix the initial value of the spatial scale parameter to be sufficiently 
large (e.g., $5$ in the simulation study and $10$ in Section \ref{real}). This is supported 
by the numerical results in \citet{Kaufman-13} that when the spatial scale parameter is 
fixed at a certain value larger than its true value, both the estimators and predictors 
can still perform well. Since the specification of the initial value for $\tau_1$ and $\tau_2$ 
almost does not affect the parameter estimation, we set their initial values to 
be certain relatively small numbers.   For the remaining parameters, we randomly select 
$100$ parameter vectors in the parameter space using Latin hypercube sampling (LHS), and 
choose the one with the smallest negative log-likelihood as the initial value. Note that 
the range of the smoothness parameters is assumed to be $(1, 5]$ since values that are too 
large would be unrealistic and result in numerical instability. From our extensive numerical 
experience, the algorithm typically converges after $100$ to $200$ iterations. Further computational gain can be achieved using 
parallel computing for gradient estimation in each iteration.

We conduct $500$ simulation runs, and generate realizations based on the cross-covariance function (\ref{gen}) with
\mjfan{\begin{equation}\label{true_value_sim}
\bm{\theta}=(1, 1, 0.5, 3, 4, 1/2, 0.1, 0.1)
\end{equation}} 
on \mjfan{three regular grids with $(n_{\rm lat}, n_{\rm lon}) = (10, 20), (15, 30)$ and $(20, 40)$, respectively.
The sample size is increased} to check the behavior of the MLEs 
from the perspective of infill asymptotics (i.e., fixed domain asymptotics), which is more 
natural than increasing domain asymptotics given a random field on a sphere. \mjfan{The regular grids are used here to speed up the computation. 
We have also conducted some smaller scale experiments on irregular grids, which give similar results (see Supplementary Materials S3).} \mjfan{Figure 
\ref{fig:sim_MLE} shows the boxplots of the MLEs. For all the parameters, the medians of the estimates are close to their true values, i.e., the estimates have small biases. Also, except for a few outliers, the small spread of the boxplots shows low variability of the estimates. \citet{Zhang-04} showed that the parameters $\sigma^2$ and $a$ of a Mat\'ern covariance function (with a fixed smoothness parameter $\nu$) in dimension $d \leq 3$ cannot be consistently estimated under infill asymptotics, while the quantity $\sigma^2a^{2\nu}$ can be consistently estimated. Since the TMM is derived from a bivariate Mat\'ern model, it is not surprising that the standard errors of the estimates for $\sigma_1$ and $\sigma_2$ do not decrease significantly when the sample size increases. To give a comprehensive picture of the accuracy 
of parameter estimation, we also conduct simulation studies for three other scenarios: varying the noise level, varying the smoothness of the field and 
including covariates in the model. The simulation results are presented in Supplementary Materials S4-S6, and these show a good performance in parameter estimation except in settings of a very high noise level (i.e., low signal-to-noise ratio)
and a very rough field that is difficult to distinguish from the observational (white) noise.}

\begin{figure}[htbp] 
   \centering
   \includegraphics[width=6in]{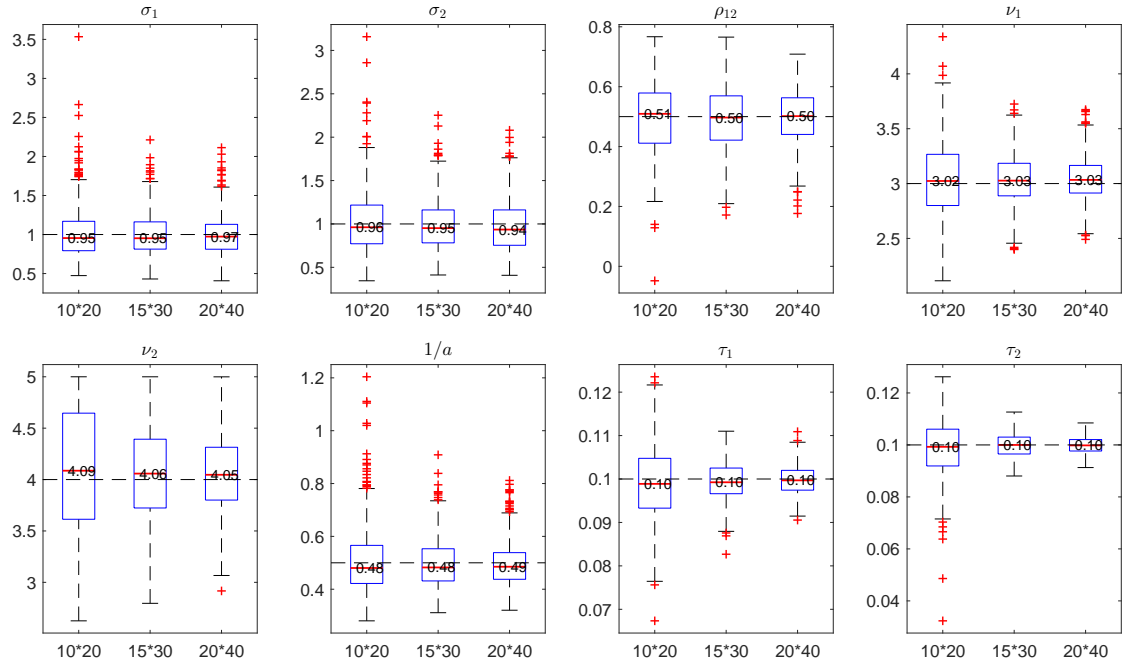}
   \caption{Results of the Monte Carlo simulation study to investigate the accuracy of 
   parameter estimation for the TMM. The MLEs of 
   \mjfan{$\bm{\theta}=(\sigma_1, \sigma_2, \rho_{12}, \nu_1, \nu_2, 1/a, \tau_1, \tau_2)$}
   are summarized by boxplots for \mjfan{three} simulated data sets \mjfan{with increasing sample size (shown on x-axis)}. 
   On each box, the central mark is the median (its value is explicitly 
   shown with two decimals), the edges of the box are the 25th and 75th percentiles, 
   the whiskers extend to the most extreme data points not considered outliers, and 
   outliers are plotted individually. The dashed horizontal lines are at the true values.}
   \label{fig:sim_MLE}
\end{figure}

\begin{figure}[htbp] 
   \centering
   \includegraphics[width=4in]{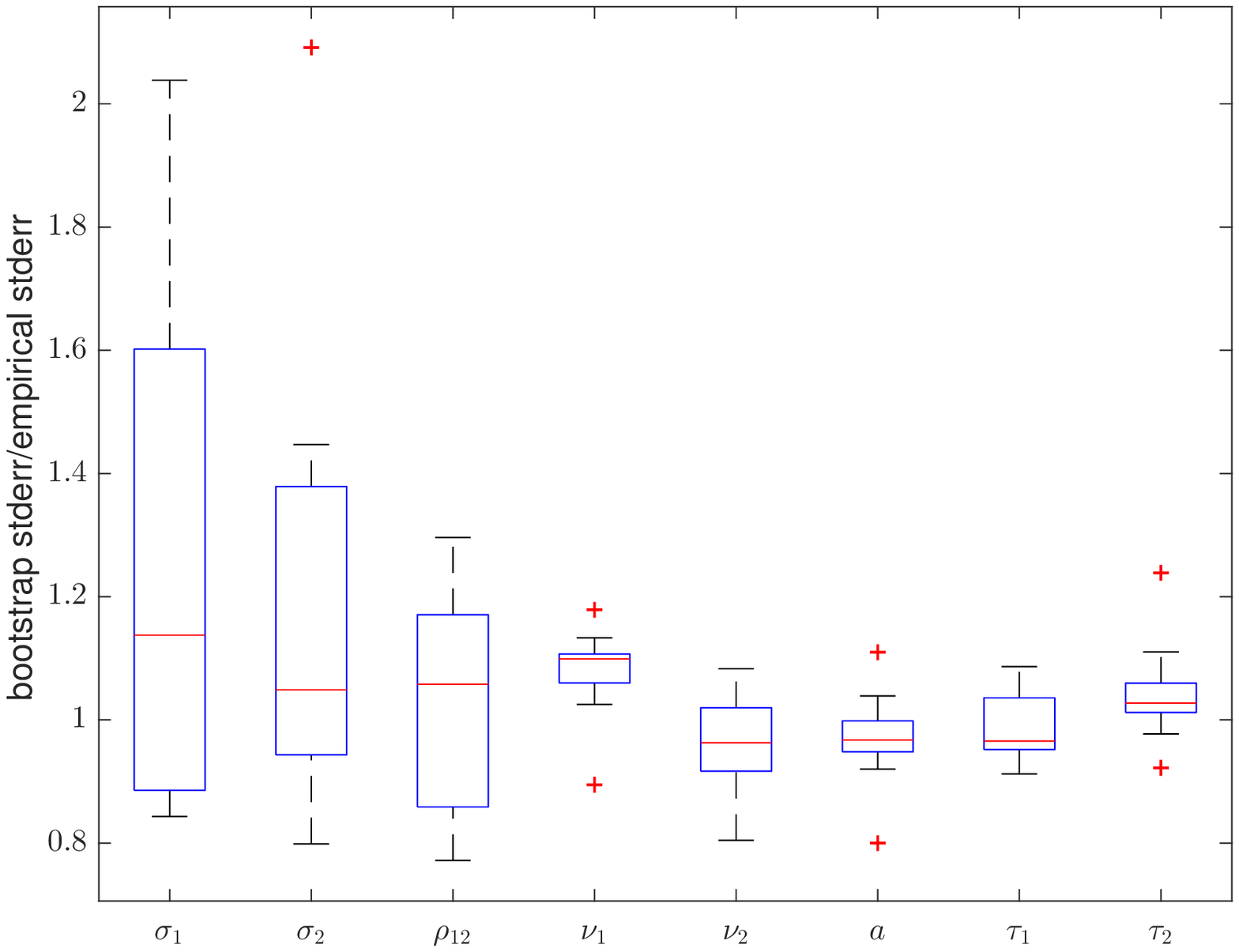}
   \caption{\mjfan{Ratios between the bootstrap and empirical standard errors. For each parameter, the 10 ratios are summarized by a boxplot.}}
   \label{fig:sim_boot}
\end{figure}

\mjfan{Asymptotic standard errors for MLEs based on the Fisher information matrix are not suitable in our context since their validity depends on the framework of increasing domain asymptotics. Therefore, we estimate the standard errors of MLEs using a parametric bootstrap procedure. 
Accordingly, the bootstrap samples are independent realizations of the TMM with parameters estimated by maximum likelihood \citep{Horowitz-01}. 
We test the  effectiveness of the parametric bootstrap on the aforementioned simulation with 
$(n_{\rm lat}, n_{\rm lon})=(15, 30)$ by comparing the empirical standard errors
and the bootstrap standard errors computed based on 200 bootstrap samples.
We obtain 10 sets of bootstrap standard errors for the first 10 of the 500 simulation runs. Figure \ref{fig:sim_boot} shows the 
ratios between the bootstrap and empirical standard errors. For each parameter, the 10 ratios are summarized by a boxplot. It is noticeable that the bootstrap standard errors tend to
be slightly higher than the empirical ones, which suggests that the former can be seen as conservative estimates of the true standard errors. 
Also, the ratios for $\sigma_1$ and $\sigma_2$ have relatively large spreads.}

\subsection{Spatial Prediction}\label{spatial_pred_sim}
\begin{figure}[htbp] 
   \centering
   \includegraphics[width=5.5in]{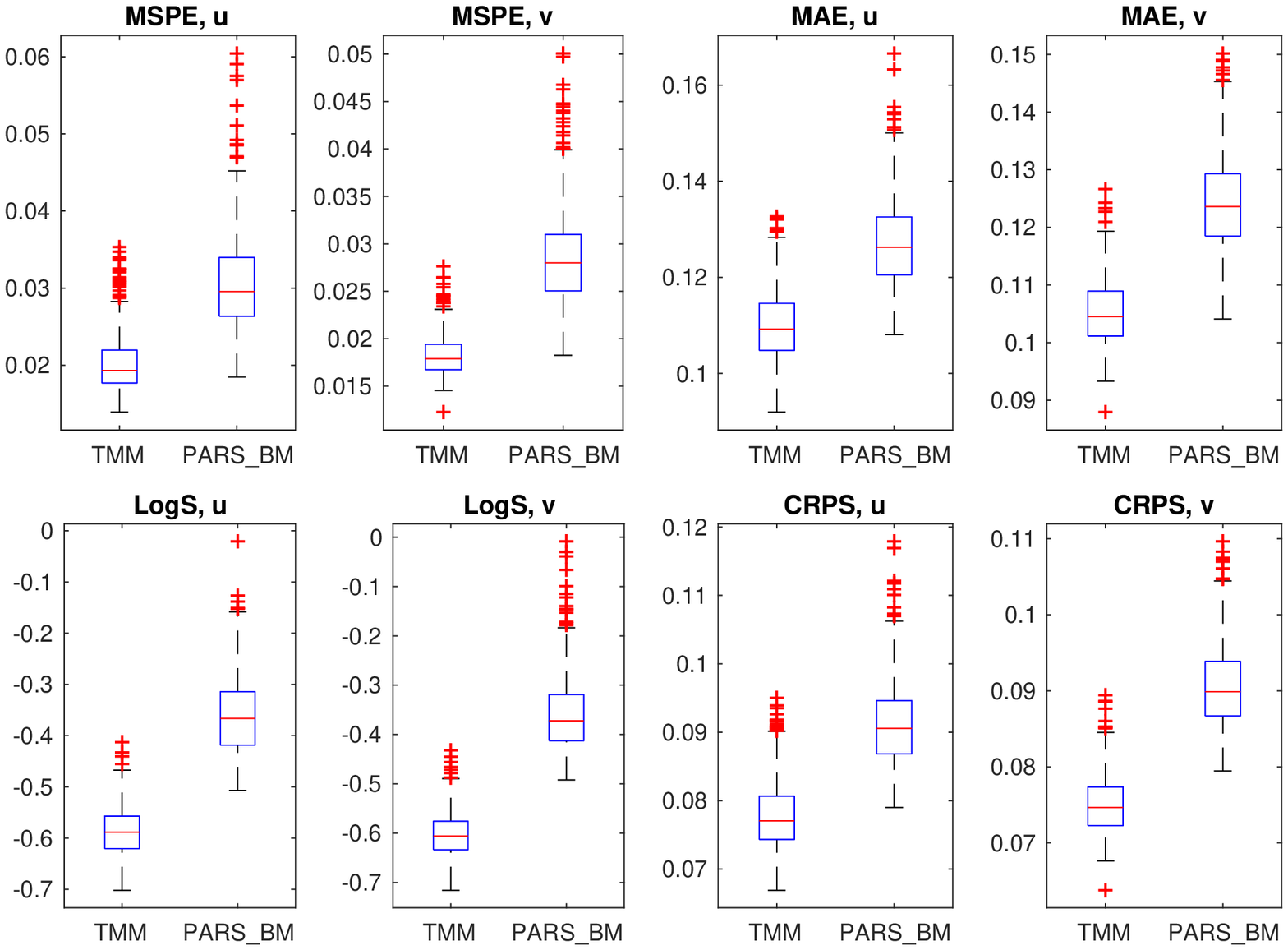}
   \caption{\mjfan{Boxplots of the four scoring rules MSPE, MAE, LogS, and CRPS for the TMM and the PARS-BM with 500 replications.}}
   \label{fig:pred_comp_sim}
\end{figure}
\mjfan{In this subsection, we compare the predictive performance of the TMM with the parsimonious bivariate Mat\'ern model (PARS-BM) (see Section \ref{para_model}) by simulation. The spatial prediction of vector fields can be achieved by cokriging; see \citet{Myers-82, Ver-93} for examples. We use the implementation of the PARS-BM in the R package RandomFields (\mjfan{Version 3.0.62}) \citep{Schlather-15}. Suppose that the data are simulated from the TMM on the HEALPix grid with 768 grid points using the same parameter specification as (\ref{true_value_sim}). Half of the locations are randomly selected outside a randomly selected longitudinal region with width $30^\circ$ for estimation, and the remaining locations are held out for prediction. In this way, both short-range and long-range predictions are taken into consideration. We assess the prediction accuracy using several scoring rules: the mean squared prediction error (MSPE), the mean absolute error (MAE), the logarithmic score (LogS) and the continuous ranked probability score (CRPS) (Gneiting and Raftery, 2007). The cross-validation procedure is repeated 500 times, and the boxplots of the four scoring rules for the two models are shown in Figure \ref{fig:pred_comp_sim}. All the boxplots of the TMM are lower than those of the PARS-BM, which indicates that the TMM has better predictive performance in terms of the four scoring rules.} 



\section{Data Example}\label{real}

In this section, we illustrate the effectiveness of the proposed TMM and the associated
statistical methodology by applying them to an ocean surface wind data set called QuikSCAT. 
Given that surface winds behave differently over 
land and ocean, here we focus on the latter. Surface wind speeds
and directions over the ocean are measured through the SeaWinds scatterometer onboard the NASA 
QuikSCAT satellite. The retrieved wind vectors produce two data products: Level 2B and Level 3, 
which are available on \url{http://podaac.jpl.nasa.gov}. The user manual \citet{Piolle-02} 
gives a comprehensive description of the data products. We use the Level 3 data set containing 
monthly mean ocean surface winds from January 2000 through December 2008. \mjfan{We focus on
modeling the horizontal component of the ocean surface winds, and thus the observations
are the zonal and meridional winds in [m/s], abbreviated as $u$ and $v$ winds hereafter. The sampling locations are on an 
(incomplete) regular grid with spatial resolution 1 degree latitude by 1 degree longitude.}

The horizontal component of a surface wind field is an important example of a naturally
occurring tangential vector field, which exhibits variability over both space and time.
Given that the spatio-temporal variability is non-separable, \citet{Cressie-99} fitted several non-separable,
spatio-temporal stationary covariance functions to a wind speed data set.
\citet[Chapter 9.4]{Cressie-11} gave a review on 
hierarchical Bayesian models for wind fields, which lead to more complicated and realistic spatio-temporal covariance structures.
\mjfan{The importance 
of representing the horizontal component of a surface wind field in terms of its curl-free and divergence-free
components can be gauged from the fact that the vorticity, which can be entirely determined by
 the divergence-free component, is a key ingredient in the analysis 
and prediction of cyclonic events \citep[Chapter 4.2]{Holton-12}.}

An extensive scientific literature \citep{Shukla-74, Bijlsma-86} suggests representing the horizontal component
of a surface wind field $\mathbf{Y}(\mathbf{s},t)$ in terms of a velocity potential $\Phi$ and a stream function $\Psi$
$$
\mathbf{Y}(\mathbf{s}, t)=\nabla_{\mathbf{s}}^*\Phi(\mathbf{s}, t)+L_{\mathbf{s}}^*\Psi(\mathbf{s}, t)
=\mathbf{P}_{\mathbf{s}}\nabla_{\mathbf{s}} \Phi(\mathbf{s}, t)+ \mathbf{Q}_{\mathbf{s}}\nabla_{\mathbf{s}} \Psi(\mathbf{s}, t),
$$
where $\mathbf{s} \in \mathbb{S}^2$, $t$ is a time index,
and $\Phi$ and $\Psi$ are scalar functions. Note that the velocity potential and stream function
are allowed to be correlated, which has been justified in \citet{Cornford-98}.
\mjfan{For the ocean surface winds,} we impose certain structures on the velocity potential and stream function, instead of
directly on the Cartesian components of the vector field. Specifically, the spatio-temporal model described
in Section \ref{spa_temp_framework} is used to model the surface wind field.

The true wind field $\mathbf{Y}$ is typically unobservable. Suppose the observations (represented
in the canonical coordinates $(\hat{\mathbf{u}}, \hat{\mathbf{v}})$ of the tangent space)
$\widetilde{\mathbf{Y}}(\mathbf{s}_i, t_j)=(\widetilde{Y}_{\rm u}(\mathbf{s}_i, t_j), \widetilde{Y}_{\rm v}(\mathbf{s}_i, t_j))^{\rm T}$, 
$i=1,\cdots, N, j=1,\cdots, T$, where $N$ is assumed to be larger than $T$,
are of the form
\begin{equation}\label{eq:observed_wind_field}
\widetilde{\mathbf{Y}}(\mathbf{s}_i, t_j) = \mathbf{T}_{\mathbf{s}}\mathbf{Y}(\mathbf{s}_i, t_j)+\bm{\epsilon}(\mathbf{s}_i, t_j),
\end{equation}
where $\bm{\epsilon}(\mathbf{s}_i, t_j)$ are observational errors that are 
modeled as i.i.d. $\mathcal{N}(\mathbf{0}, \mathrm{diag}(\tau_1^2, \tau_2^2))$.
Due to limitations of space, and in the interest of a focused analysis, 
our goal here is primarily to demonstrate the effectiveness of the proposed model in
describing the zonal ($u$) and meridional ($v$) components of the small-scale component of 
the surface wind field, than to conduct a detailed statistical analysis of the 
full-scale surface wind field. We subtract a crude estimate of the large-scale component from
the surface wind field as a way of extracting the small-scale component. The large-scale
component of the surface wind field is estimated from the data by the vector empirical orthogonal 
function (VEOF) method \citep{Pan-01, Pan-03}. 
Let $\widetilde{\mathbf{Y}}= (\widetilde{\mathbf{Y}}_{\rm u}, \widetilde{\mathbf{Y}}_{\rm v})$ denote the 
matrix of observations,  where $\widetilde{\mathbf{Y}}_{\rm u}$ and $\widetilde{\mathbf{Y}}_{\rm v}$ denote the $T\times N$ 
matrices corresponding to the $u$ and $v$ components, respectively, of the observed surface wind fields. 
We center $\widetilde{\mathbf{Y}}$ by subtracting the vector 
of column averages from each row. The singular value decomposition (SVD) is applied to $\widetilde{\mathbf{Y}}$
$$
(\widetilde{\mathbf{Y}}_{\rm u}, \widetilde{\mathbf{Y}}_{\rm v})=
\widetilde{\mathbf{Y}}=\mathbf{U}\mathbf{D}\mathbf{W}^{\rm T} = \mathbf{U}\mathbf{D}(\mathbf{W}^{\rm T}_{\rm u}, \mathbf{W}^{\rm T}_{\rm v}),
$$
where $\mathbf{U}$ is a $T\times T$ orthogonal matrix with $u_k(t_j)$ in row $j$, column $k$, 
$\mathbf{D}=\mathrm{diag}(d_1, \cdots, d_T)$ with singular values $d_1\geq d_2 \geq \cdots \geq d_T\geq 0$, and $\mathbf{W}$ is a $2N\times T$ matrix with orthonormal columns, which are VEOFs.
Element-wisely, we can express the observations as
$$
\widetilde{\mathbf{Y}}(\mathbf{s}_i, t_j)=\sum \limits_{k=1}^T d_k u_k(t_j)(w_{\mathrm{u}, k}(\mathbf{s}_i), w_{\mathrm{v}, k}(\mathbf{s}_i))^{\rm T},
$$
where $w_{\mathrm{u}, k}(\mathbf{s}_i), w_{\mathrm{v}, k}(\mathbf{s}_i)$ are the elements of $\mathbf{W}_{\rm u}$ and 
$\mathbf{W}_{\rm v}$ in row $i$, column $k$, respectively.
The cumulative sum of the ordered squared singular values suggests that the first 64 
VEOFs explain approximately 95\% of the variability in the observed surface wind fields. \mjfan{It is not surprising to see
$64$ VEOFs subtracted since the VEOF method is applied to the $u$ and $v$ components simultaneously over the whole globe.} We use this as a crude measure 
of the large-scale variability in the data, and thus approximate the large-scale component by
$$
\sum \limits_{k=1}^Kd_ku_k(t_j)(w_{\mathrm{u}, k}(\mathbf{s}_i), w_{\mathrm{v}, k}(\mathbf{s}_i))^{\rm T},
$$
where $K=64$. The residuals after subtracting the large-scale component are regarded as the 
small-scale component, which are corrupted by observational error.  The sample autocorrelation function supports our 
assumption that the residuals are temporally uncorrelated.


\begin{figure}[htbp] 
   \centering
   \includegraphics[width=4.5in]{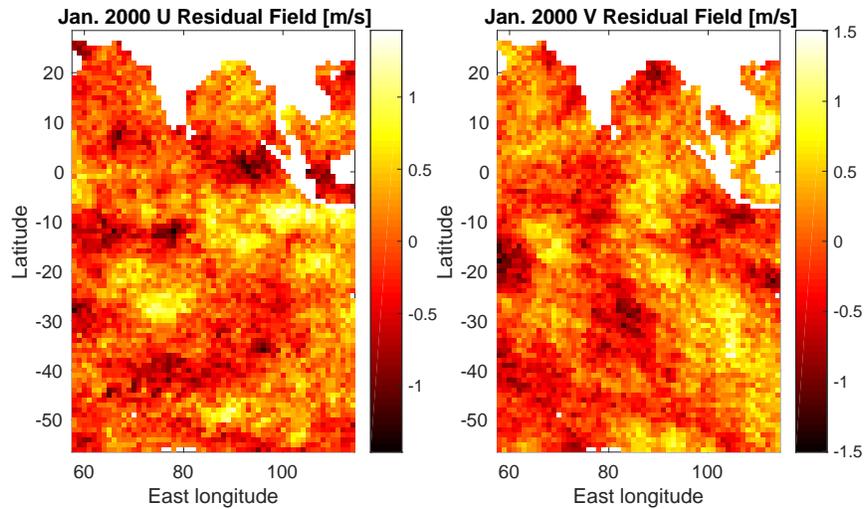}
   \caption{An example of the $u$ and $v$ residual wind fields for January, 2000 in the subregion of the Indian Ocean after subtracting the first $64$ VEOFs.}
   \label{fig:real_res_uv1}
\end{figure}

To reduce computational burden and for a quick comparison among different models, we choose a subregion of the 
Indian Ocean ($57.30$\textdegree E$-114.59$\textdegree E, $57.30$\textdegree S$-28.65$\textdegree N) with 
spatial resolution 2 degree latitude by 2 degree longitude. The subregion is large enough 
(the range of latitudes is almost $90^\circ$) so that models that treat the domain as a subset of the Euclidean 
plane cannot handle the distortion caused by the curvature. The effect of the curvature on vector fields 
is more distinct than that on scalar fields. Using a rougher grid may lead to loss of information, 
but the parameter estimates are comparable to those on the full grid 
according to our numerical experience. \mjfan{Note that there are $1070$ observations for each month,  thus $108 \times 1070$ observations in total, which are assumed to be independent across months.}

\begin{figure}[htbp] 
   \centering
   \includegraphics[width=5in]{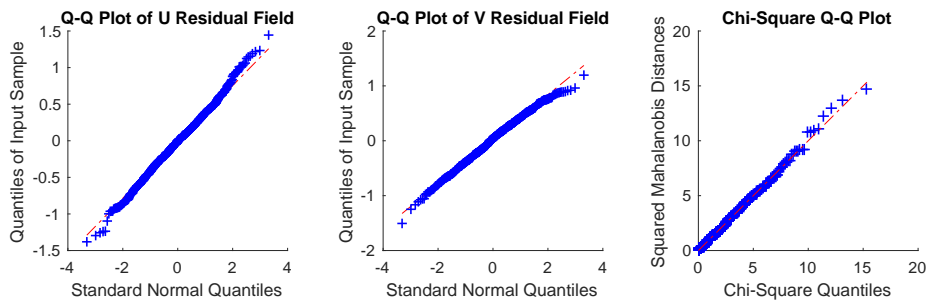}
   \caption{Marginal Q-Q plots (the first two) and Chi-Square Q-Q plot (the third) of the $u$ and $v$ residual wind fields for January, 2000 in the subregion of the Indian Ocean.}
   \label{fig:check_normality}
\end{figure}

Figure \ref{fig:real_res_uv1} contains an example of the $u$ and $v$ components of the residual wind field 
for January, 2000 in the subregion of the Indian Ocean. The marginal Q-Q plots and the 
Chi-Square Q-Q plot, shown in Figure \ref{fig:check_normality} for January, 2000, suggest that the Gaussianity assumption for the $u$ and $v$ residual wind fields is
reasonable.


We fit the TMM to the residual wind fields for all the months by minimizing the negative joint 
log-likelihood function. \mjfan{Without loss of generality, the Earth is treated as a unit sphere.} The MLEs and their bootstrap standard errors 
are listed in the second column of Table \ref{MLEs}. \mjfan{The bootstrap standard errors are calculated using the parametric bootstrap
introduced in Section \ref{accu} with $200$ bootstrap samples. Notice that the estimated standard deviation of the divergence-free component is almost twice of that of the curl-free component. 
This suggests that the divergence-free component is the dominant component of the ocean surface residual wind fields. In fact, a purely divergence-free wind field, called geostrophic wind, is the theoretical wind that results from an exact force balance between the Coriolis effect and the pressure gradient force. Surface winds that blow over the ocean are close to being geostrophic because of the relatively smooth ocean surface \citep[page 283]{Park-01}. Thus, the two standard deviation parameters, which measure the magnitude of the curl-free and the divergence-free components, carry important geophysical meanings. Moreover, }the estimated co-located correlation coefficient is significantly 
different from zero, which justifies the necessity of allowing the curl-free and the divergence-free components 
to be correlated. The signal-to-noise ratios of the $u$ and $v$ residual wind fields at $\mathbf{s}$ are
$\mbox{Var}(u(\mathbf{s}))/\tau_1^2=4.040$ and $\mbox{Var}(v(\mathbf{s}))/\tau_2^2=4.653$, respectively, which 
do not vary with respect to the location due to Proposition \ref{prop1}.

\begin{table}[htp]
\caption{Maximum likelihood estimates of the parameters for the TMM and the PARS-BM applied to the $u$ and $v$ residual wind fields in the subregion of the Indian Ocean from January 2000 through December 2008}
\begin{center}
\begin{tabular}{ccc}
\hline
\hline
Model & PARS-BM & TMM \\
\hline
$\sigma_1$ & 0.429 (3.14e-3) & 0.029 (4.25e-4)\\
$\sigma_2$ & 0.396 (3.09e-3)& 0.055 (8.05e-4)\\
$\rho_{12}$ & -0.080 (6.88e-3) & 0.281 (7.05e-3)\\
$\nu_1$ & 1.239 (0.038) & 1.758 (0.022)\\
$\nu_2$ & 1.132 (0.035) &  2.034 (0.020)\\
$1/a$ & 0.058 (1.45e-3) & 0.106 (1.80e-3) \\
$\tau_1$ & 0.218 (1.37e-3) & 0.210 (1.48e-3)\\
$\tau_2$ & 0.203 (1.61e-3) & 0.196 (1.48e-3)\\
\hline
Log-likelihood & -46995 & -45126\\
\# of parameters & 8 & 8\\
\hline
\end{tabular}
\end{center}
\label{MLEs}
\end{table}%

\begin{figure}[htbp] 
   \centering
   \includegraphics[width=5in]{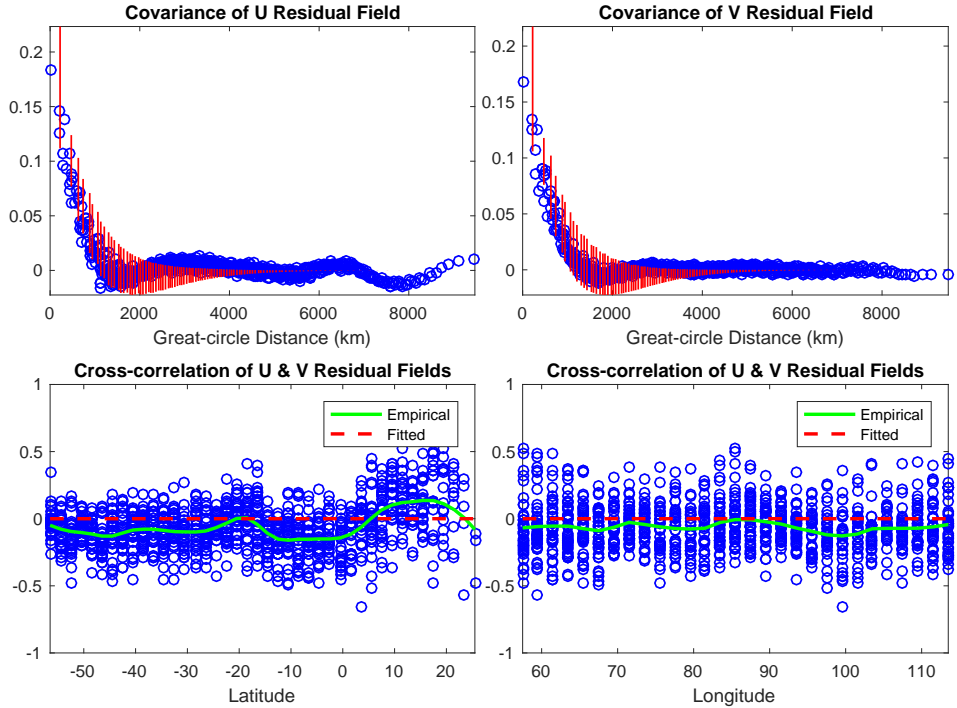}
   \caption{First row: Empirical and fitted covariances of the $u$ and $v$ residual fields. Bin medians of the empirical covariances are shown by circles. The spread (min to max) of the fitted covariances (due to \mjfan{anisotropy}) at each great-circle distance is summarized as \mjfan{a vertical line}. Second row: Empirical (circles) and fitted (dashed line) co-located cross-correlations of the $u$ and $v$ residual fields,  i.e., $\mbox{Corr}\left(u(\mathbf{s}), v(\mathbf{s})\right)$. They are plotted with respect to latitude (left) and longitude (right). The solid line represents the \mjfan{loess curve fitted to the empirical co-located cross-correlations}. }
   \label{fig:emp_fitted}
\end{figure}

\begin{figure}[htbp] 
   \centering
   \includegraphics[width=5in]{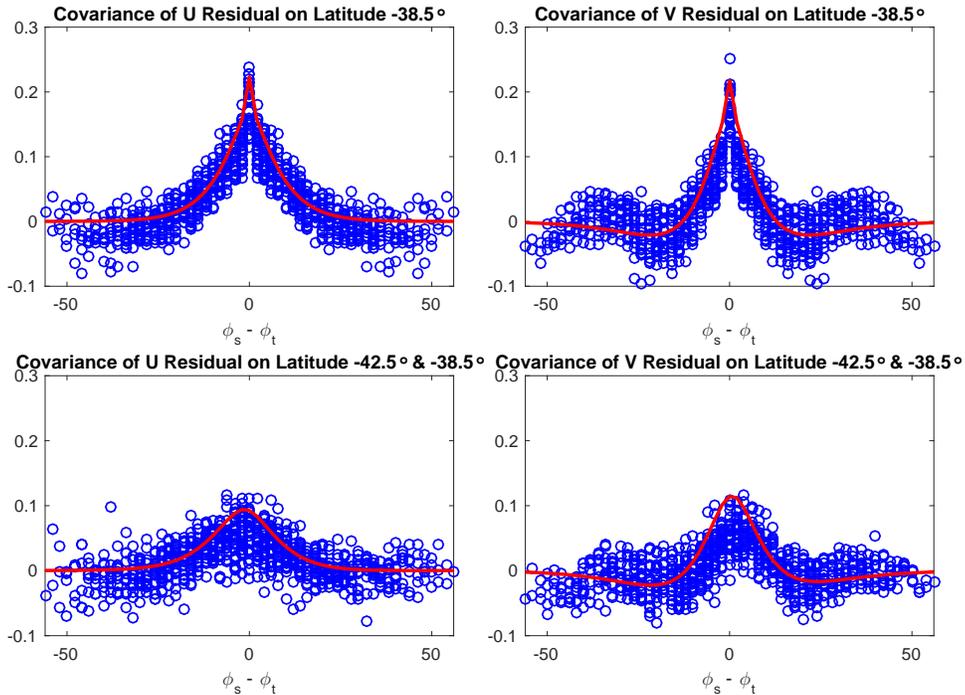}
   \caption{Empirical (circles) and fitted (solid line) covariances of the $u$ and $v$ residual fields. The covariances can be 
   represented as a function of $\theta_{\textbf{s}}$, $\theta_{\textbf{t}}$ and $\phi_{\textbf{s}}-\phi_{\textbf{t}}$. First row: $\theta_{\textbf{s}}=\theta_{\textbf{t}}=-38.5^{\circ}$; second row: $\theta_{\textbf{s}}=-42.5^{\circ}$, $\theta_{\textbf{t}}=-38.5^{\circ}$.}
   \label{fig:emp_fitted_cov_u_v}
\end{figure}

\begin{figure}[htbp] 
   \centering
   \includegraphics[width=5in]{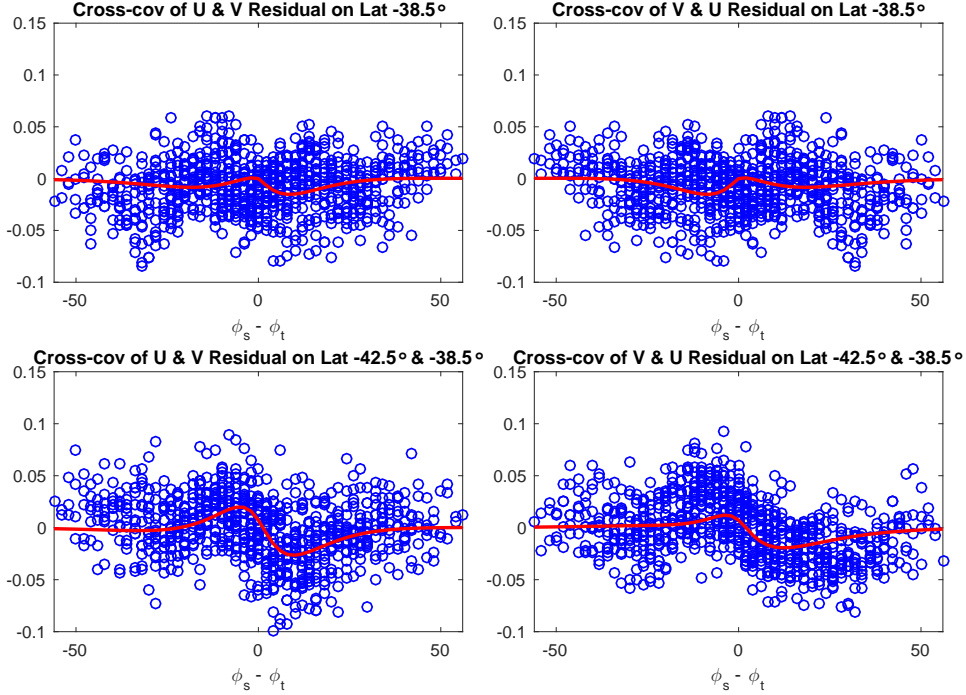}
   \caption{Empirical (circles) and fitted (solid line) cross-covariances of the $u$ and $v$ residual fields. First column: $\mbox{Cov}\left(u(\textbf{s}), v(\textbf{t})\right)$; second column:  $\mbox{Cov}\left(v(\textbf{s}), u(\textbf{t})\right)$. The cross-covariances can be 
   represented as a function of $\theta_{\textbf{s}}$, $\theta_{\textbf{t}}$ and $\phi_{\textbf{s}}-\phi_{\textbf{t}}$. First row: $\theta_{\textbf{s}}=\theta_{\textbf{t}}=-38.5^{\circ}$; second row: $\theta_{\textbf{s}}=-42.5^{\circ}$, $\theta_{\textbf{t}}=-38.5^{\circ}$.}
   \label{fig:emp_fitted_cross_cov_u_v}
\end{figure}

Figure \ref{fig:emp_fitted} shows the empirical and fitted covariances and co-located cross-correlations, i.e., $\mbox{Corr}\left(u(\mathbf{s}), v(\mathbf{s})\right)$, of the $u$ and $v$ residual fields. The fitted covariances are \mjfan{generally} in agreement with the empirical ones. In particular, as a well-accepted characteristic in meteorological variables \citep[Chapter 4.3]{Daley-91}, the negative covariances around the great-circle distance of $2000$ km are captured by the fitted model. According to Proposition \ref{prop1}, the fitted co-located cross-correlations under the TMM are identically zero. \mjfan{On the other hand, we observe that the empirical co-located cross-correlations are nearly zero except for a mild oscillation around zero, which can be attributed to the subtraction of large-scale components. Therefore, the observations seem to conform to the 
proposed model in terms of this important characteristic.} \mjfan{Since the TMM assumes that the vector field is axially symmetric, in Figure \ref{fig:emp_fitted_cov_u_v} we plot the empirical and fitted covariances of the $u$ and $v$ residual fields as a function of $\theta_{\textbf{s}}$, $\theta_{\textbf{t}}$ and $\phi_{\textbf{s}}-\phi_{\textbf{t}}$ for certain latitudes. For the same reason, Figure \ref{fig:emp_fitted_cross_cov_u_v} shows the empirical and fitted cross-covariances of the $u$ and $v$ residual fields for certain latitudes. The fitted model captures the peak of the empirical covariances around $\phi_{\textbf{s}}=\phi_{\textbf{t}}$. The trend of the empirical cross-covariances, which is asymmetric with respect to the point of  $\phi_{\textbf{s}}=\phi_{\textbf{t}}$, is also captured by the fitted model. The empirical and fitted variances of the $u$ and $v$ residual fields match well with each other, as shown in Supplementary Materials S7.}

\begin{table}[htp]
\caption{Cokriging cross-validation scores averaged over $20$ cross-validation replications for the two models on the $u$ and $v$ residual wind fields. For each case, the standard deviation of the scores over $20$ replications is shown in parenthesis}
\begin{center}
\begin{tabular}{ccccccccc}
\hline
\hline
Model & Variable & MSPE & MAE & LogS & CRPS \\
\hline
\multirow{2}{*}{PARS-BM} & $u$ & 0.1191 (0.0020) & 0.2650 (0.0022) & 0.3153 (0.0079) & 0.1891 (0.0015)\\
& $v$ & 0.1062 (0.0015) & 0.2531 (0.0016) & 0.2595 (0.0074) & 0.1795 (0.0012)\\
\hline
\multirow{2}{*}{TMM} & $u$ & 0.1126 (0.0020) & 0.2587 (0.0021) & 0.2933 (0.0078) & 0.1845 (0.0015)\\
& $v$ & 0.1034 (0.0017) & 0.2495 (0.0020) & 0.2484 (0.0084) & 0.1772 (0.0014)\\
\hline
\end{tabular}
\end{center}
\label{cokriging}
\end{table}%

For comparison, \mjfan{we also fit the parsimonious bivariate Mat\'ern model (PARS-BM) by the R package RandomFields}. The results are given in the first column of Table \ref{MLEs}, where the standard errors are calculated \mjfan{using the parametric bootstrap (see Section \ref{accu})}. With the same number of parameters, the PARS-BM has a lower value of the log-likelihood. 


Apart from model fitting, we further compare the predictive performance of the two models. 
To estimate the parameters, we randomly select half of the $1070$ locations outside a $20^\circ$ (width) by $40^\circ$ (height) 
rectangular region in the center of the subregion of the Indian Ocean. The selected 
locations for parameter estimation are consistent over all time points. To predict the 
values at the remaining locations using the parameter estimates just obtained, 
cokriging is performed for each time point. \mjfan{The same scoring rules as in Section \ref{spatial_pred_sim}} are used to assess 
the prediction accuracy, averaged over all the predicted locations and time points.  
We repeat the cross-validation procedure $20$ times, and calculate the mean and standard deviation 
(in parenthesis) of the resulting scores, which are displayed in Table \ref{cokriging}. 
The average scores of the TMM are all smaller than the PARS-BM. A series 
of two-sample $t$-tests for each pair of the average scores of the two models suggest that 
the TMM \mjfan{statistically outperforms} the PARS-BM ($p$-value $< 0.05$ in each case).


\begin{figure}[htbp] 
   \centering
   \includegraphics[width=5in]{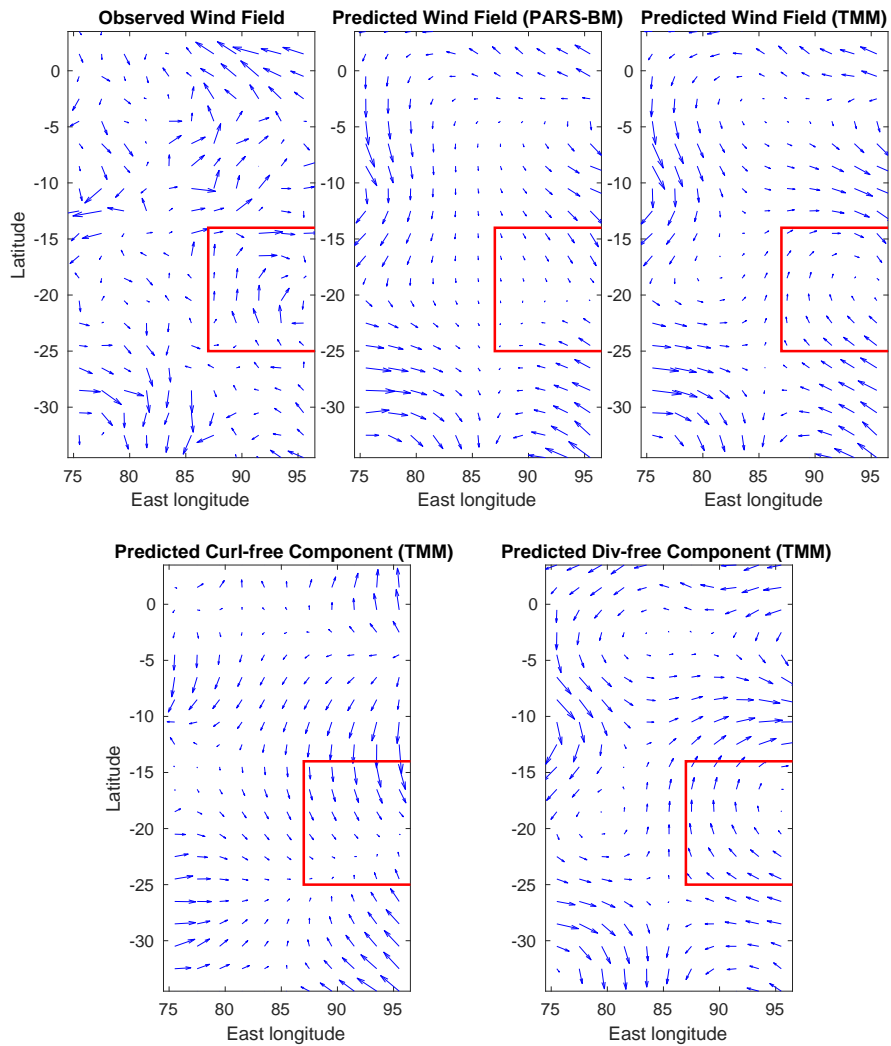}
   \caption{Observed and predicted wind fields by the TMM and the PARS-BM within the rectangular region for January, 2000. 
   \mjfan{The second row shows the predicted curl-free and divergence-free components of the wind field by the TMM. 
   Note that the arrows have been automatically scaled to fit within the grid.}}
   \label{fig:comp}
\end{figure}

Finally, we compare the wind fields predicted by the two models with the observed one. 
The former are obtained by cokriging, which uses the locations outside the rectangular region to predict those 
within it, while the parameters are estimated based on the entire data set (see Table \ref{MLEs}). Figure \ref{fig:comp} 
shows the comparison among the wind fields within the rectangular region for January, 2000. 
It is not surprising that there are some discrepancies between the predicted and observed 
wind fields. 
Nonetheless, the predicted wind field by the 
TMM captures the rotational motion within the marked rectangle, which is missed by the PARS-BM.

\section{Discussion}\label{dis}

\mjfan{We have introduced a general framework to construct models for tangential vector fields on the unit sphere, using the surface gradient or the surface curl operator. Under this framework, we have presented a new class of parametric models, named Tangent Mat\'ern Model, which is derived by using a bivariate Mat\'ern model as the underlying potential field. Unlike most parametric models for vector fields that do not impose specific physical constraints, our model is directly motivated by the celebrated Helmholtz-Hodge decomposition, in which a tangential vector field is naturally decomposed into the sum of a curl-free and a divergence-free components. These two components carry important scientific meanings for describing the phenomena 
within the context of naturally occurring vector fields such
as surface wind fields, oceanic currents, etc.}

\mjfan{We have compared the Tangent Mat\'ern Model with the parsimonious bivariate Mat\'ern model on a QuikSCAT ocean surface wind velocity data set in terms of model fitting and spatial prediction. We focused on modeling the small-scale component, where the large-scale spatio-temporal component was subtracted using empirical orthogonal functions. In the comparison, our model captured the negative covariances and the rotational flow of the wind field, while the parsimonious bivariate Mat\'ern model did not. As a result, the prediction error produced by our model was significantly smaller than that by the parsimonious bivariate Mat\'ern model. This demonstrates the importance of incorporating these physical characteristics into the models for terrestrial physical processes.}

\mjfan{A limitation of the Tangent Mat\'ern Model is the assumption of isotropy for the underlying potential field. This isotropy assumption leads to Proposition \ref{prop1}, in which the co-located cross-covariance function is constant over the sphere, and in particular,  $\mbox{Corr}\left(u(\textbf{s}), v(\textbf{s})\right)=0$. Nonetheless, we can extend the model using an anisotropic underlying potential field, such as non-stationary bivariate Mat\'ern models with spatially varying parameters \citep{Kleiber-12, Jun-14}. The spatially varying parameters can be modeled by parametric functions of spherical harmonics \citep{Bolin-11} or covariates observed together with the vector field \citep{Risser-15}. This remains a topic of future research.}

\mjfan{Proposition \ref{prop2} reveals the connection of our model with the models of \citet{Jun-11, Jun-14, JunS2008}, although they are derived from different perspectives. Also, as demonstrated by Proposition \ref{prop2}, the tangential vector field that follows our model is axially symmetric when represented in the spherical coordinate system. Thus, this result enables fast computation for large data sets when the observations are on a regular grid. For example, on average, it takes approximately $8$ hours (on a machine with a 2.60GHz Intel Xeon E5-2690 v3 processor) to estimate the parameters when the sample size is as large as $5000$. For irregularly spaced observations, apart from the method of covariance tapering mentioned in Section \ref{para_est}, a multi-resolution model constructed by the Wendland radial basis functions can also be used to represent or approximate the underlying potential field with a Mat\'ern or a more complicated covariance structure \citep{Nychka-14}. Since the key matrices involved in the computation are sparse, this model can be applied to large data sets.}

\section*{Appendix}
\textbf{Appendix A: Notation and Definition}

\vspace{0.5cm}

\mjfan{In this subsection, we describe some auxiliary notations and definitions that are helpful for understanding the construction
and characteristics of tangential vector fields on the unit sphere.}
\begin{equation}\label{eq:Q_s}
\mathbf{Q}_{\mathbf{s}}=: \hat{\mathbf{r}} \times \mathbf{P}_{\mathbf{s}}=\left( \begin{matrix}
0 & -s_3 & s_2\\
s_3 & 0 & -s_1\\
-s_2 & s_1 & 0
\end{matrix} \right).
\end{equation}

\mjfan{The surface gradient and the surface curl operators
can be represented in the spherical coordinate system except at the two poles,} 
$$
\nabla_{\mathbf{s}}^{*}f=\frac{\partial f}{\partial \theta}\hat{\bm{\theta}}
+ \frac{1}{\sin\theta}\frac{\partial f}{\partial \phi}\hat{\bm{\phi}},
$$
and
$$
L_{\mathbf{s}}^{*}f=\frac{\partial f}{\partial \theta}\hat{\bm{\phi}}-\frac{1}
{\sin\theta}\frac{\partial f}{\partial \phi}\hat{\bm{\theta}}.
$$
Let $X(\mathbf{s})$ denote a scalar random field on a spherical shell $S_{\epsilon}$, where $(\Omega, \mathcal{F}, \mathbb{P})$
is the underlying probability space. The partial derivative of $X(\mathbf{s})$ in quadratic mean along
the $i$-th coordinate direction is defined as a random process
$D_{\mathrm{qm}}^{(i)}X(\mathbf{s})$ such that
$$
\mathbb{E}\left(\frac{X(\mathbf{s}+h\mathbf{e}_i)-X(\mathbf{s})}{h}-D_{\mathrm{qm}}^{(i)}X(\mathbf{s})\right)^2\rightarrow 0
\qquad \mbox{as}~h\rightarrow 0,
$$
where $\mathbf{e}_i$ is the unit vector in $\mathbb{R}^3$ with the $i$-th element one and zeros elsewhere.
The sample partial derivatives of $X(\mathbf{s})$, $D^{(i)}X(\mathbf{s}), i=1,2,3$, are defined as the usual
partial derivatives of $X(\mathbf{s}, \omega)$ for any fixed $\omega \in \Omega$. A random field $\widetilde{X}$ is said to be a $\mathbb{P}-$a.e. sample continuously differentiable version of $X$ if the following hold:
\begin{itemize}
\item[(i)] For any $\mathbf{s} \in S_{\epsilon}$, $\mathbb{P}\left(\widetilde{X}(\mathbf{s})=X(\mathbf{s})\right)=1$.
\item[(ii)] There exists $\Omega_0 \subset \Omega$ with $\mathbb{P}(\Omega_0)=1$
such that $\widetilde{X}(\mathbf{s}, \omega)$ is continuously differentiable on $S_{\epsilon}$
for any $\omega \in \Omega_0$.
\end{itemize}
A random tangential vector field on $\mathbb{S}^2$ is said to be curl-free or divergence-free
if its sample paths are curl-free or divergence-free $\mathbb{P}-$a.e.

\vspace{1cm}

\noindent\textbf{Appendix B: Proof of Theorem \ref{main_thm}}

\vspace{0.5cm}

We first verify (\ref{curl_cov}), and then (\ref{div_cov}) can be derived similarly. Since $D^{(i)}\widetilde{Z}(\mathbf{s})=D_{\mathrm{qm}}^{(i)}Z(\mathbf{s})$ $\mathbb{P}-$a.e., $\mathbf{Y}_{\mathrm{curl}, Z}(\mathbf{s})=\mathbf{P}_{\mathbf{s}}\nabla_{\mathbf{s}, \mathrm{qm}}Z(\mathbf{s})$ $\mathbb{P}-$a.e.,
where $\nabla_{\mathbf{s}, \mathrm{qm}}$ is the usual gradient on $\mathbb{R}^3$ defined in the sense of quadratic mean. Thus,
\begin{eqnarray*}
\mathbf{C}_{\mathrm{curl}, Z}(\mathbf{s}, \mathbf{t})
&=&
\mathbf{P}_{\mathbf{s}}\mathrm{Cov}\left(\nabla_{\mathbf{s}, \mathrm{qm}}Z(\mathbf{s}), \nabla_{\mathbf{t}, 
\mathrm{qm}}Z(\mathbf{t})\right)\mathbf{P}_{\mathbf{t}}^{\rm T}\\
&=&
\mathbf{P}_{\mathbf{s}} \left[\mathrm{Cov}\left(D_{\mathrm{qm}}^{(i)}Z(\mathbf{s}), 
D_{\mathrm{qm}}^{(j)}Z(\mathbf{t}) \right)\right]_{1\leq i, j \leq 3}\mathbf{P}_{\mathbf{t}}^{\rm T}\\
&=&
\mathbf{P}_{\mathbf{s}} \left[\frac{\partial^2 C}{\partial s_i \partial t_j}(\mathbf{s}-\mathbf{t})\right]_{1\leq i, j \leq 3}
\mathbf{P}_{\mathbf{t}}^{\rm T} \qquad \mbox{by condition \textbf{A2}}\\
&=& 
-\mathbf{P}_{\mathbf{s}} \left[\frac{\partial^2 C}{\partial h_i \partial h_j}(\mathbf{h})\right]_{1\leq i, j \leq 3}\bigg{|}_{\mathbf{h}=\mathbf{s}-\mathbf{t}}\mathbf{P}_{\mathbf{t}}^{\rm T} \\
&=&-\mathbf{P}_{\mathbf{s}}\nabla_{\mathbf{h}}\nabla_{\mathbf{h}}^{\rm T}C(\mathbf{h}) \bigg{|}_{\mathbf{h}=\mathbf{s}-\mathbf{t}} \mathbf{P}_{\mathbf{t}}^{\rm T}.
\end{eqnarray*}

\vspace{1cm}

\noindent\textbf{Appendix C: Proof that $\nu>1$ is necessary and sufficient}

\vspace{0.5cm}

First, $\nu>1$ is necessary since the underlying potential field $Z(\mathbf{s})$ is required 
to be differentiable in quadratic mean. Second, we shall show the sufficiency. With $\nu>1$ 
and its Gaussianity,  the sample paths of $Z$ are continuously differentiable $\mathbb{P}-$a.e. 
\citep{Handcock-94}. Moreover, $Z$ has partial derivatives in quadratic mean. They imply 
that $\left(Z(\mathbf{s}+h\mathbf{e}_i)-Z(\mathbf{s})\right)/h$ converges to $D_{\mathrm{qm}}^{(i)}Z(\mathbf{s})$ 
in $L_2$ (and in probability), and converges to $D^{(i)}Z(\mathbf{s})$ almost everywhere 
(and in probability). By the uniqueness of convergence in probability, 
$D_{\mathrm{qm}}^{(i)}Z(\mathbf{s})=D^{(i)}Z(\mathbf{s})$ $\mathbb{P}-$a.e. Thus, we can choose 
$Z$ as $\widetilde{Z}$ so that condition \textbf{A1} is satisfied. \citet{JunS2007} pointed out 
that the Mat\'ern model (\ref{matern}) is twice continuously differentiable and 
$M'(0; \nu, a)=0$. Then $C_1(\lVert \mathbf{h} \rVert)$ is also twice continuously differentiable 
when $\lVert \mathbf{h} \rVert>0$. It can be shown by definition that
$$
\frac{\partial C_1(\lVert \mathbf{h} \rVert)}{\partial h_i}
=\begin{cases}
C'_1(\lVert \mathbf{h} \rVert)\frac{h_i}{\lVert \mathbf{h} \rVert} & \mbox{if } \mathbf{h}\neq \mathbf{0}\\
0 & \mbox{if } \mathbf{h}=\mathbf{0}.
\end{cases}
$$
Based on the first derivatives, we can also show by definition that
$$
\frac{\partial^2C_1(\lVert \mathbf{h} \rVert)}{\partial h_i \partial h_j}\bigg|_{\mathbf{h}=\mathbf{0}}
=\begin{cases}
0 & \mbox{if } i\neq j\\
C_1''(0) & \mbox{if } i=j.
\end{cases}
$$
This is equivalent to
$$
\nabla_{\mathbf{h}} \nabla_{\mathbf{h}}^{\rm T}C_1(\lVert \mathbf{h} \rVert)\bigg|_{\mathbf{h}=\mathbf{0}}=F(0)\mathbf{I}_3.
$$
Together with
$$
\lim_{\mathbf{h} \to \mathbf{0}}\nabla_{\mathbf{h}} \nabla_{\mathbf{h}}^{\rm T}C_1(\lVert \mathbf{h} \rVert)\bigg|_{\mathbf{h}\neq \mathbf{0}}
=\lim_{\mathbf{h} \to \mathbf{0}} F(\lVert \mathbf{h} \rVert)\mathbf{I}_3+G(\lVert \mathbf{h} \rVert)\mathbf{h}\mathbf{h}^{\rm T}=F(0)\mathbf{I}_3,
$$
it implies that $C_1(\lVert \mathbf{h} \rVert)$ is twice continuously differentiable at 
$\mathbf{h}=\mathbf{0}$. Thus, condition \textbf{A2} is satisfied.

\vspace{1cm}

\noindent\textbf{Appendix D: Isotropic $Z(\mathbf{s})$}

\vspace{0.5cm}

Since $C_1(\lVert \mathbf{h} \rVert)$ is twice continuously differentiable
when $h_1\geq 0$ and $h_2=h_3=0$, $C_1(r)$ is also twice continuously differentiable. Moreover, the fact that
$C_1(\lVert \mathbf{h} \rVert)$ is continuously differentiable at $\mathbf{h}=\mathbf{0}$ implies that $C_1'(0)=0$. Then we have
\begin{equation}\label{mat_K}
\nabla_{\mathbf{h}} \nabla_{\mathbf{h}}^{\rm T}C_1(\lVert \mathbf{h} \rVert)
=\begin{cases} F(\lVert \mathbf{h} \rVert)\mathbf{I}_3+G(\lVert \mathbf{h} \rVert)\mathbf{h}\mathbf{h}^{\rm T} & \mbox{if } \mathbf{h}\neq \mathbf{0} \\
F(0)\mathbf{I}_3 & \mbox{if } \mathbf{h}=\mathbf{0},
\end{cases}
\end{equation}
where
\begin{equation}\label{fun_F_gen}
F(r)=\begin{cases} \frac{1}{r}C_1'(r) & \mbox{if } r>0\\
C_1''(0) & \mbox{if } r=0,
\end{cases}
\end{equation}
and
\begin{equation}\label{fun_G_gen}
G(r)=\frac{1}{r}\left( \frac{1}{r}C_1'(r) \right)' \quad \mbox{if } r>0.
\end{equation}
Here the value of $\nabla_{\mathbf{h}} \nabla_{\mathbf{h}}^{\rm T}C_1(\lVert \mathbf{h} \rVert)$ at $\mathbf{h}=\mathbf{0}$ is obtained by
$$
\lim_{\mathbf{h} \to \mathbf{0}} F(\lVert \mathbf{h} \rVert)\mathbf{I}_3+G(\lVert \mathbf{h} \rVert)\mathbf{h}\mathbf{h}^{\rm T}=F(0)\mathbf{I}_3.
$$
\mjfan{When $C_1$ is chosen as the Mat\'ern model, we have the explicit expressions of $F$ and $G$
as}
\begin{equation}\label{fun_F}
F_{\textrm{Mat}}(r; \nu, a)=\begin{cases} -\frac{2^{1-\nu}}{\Gamma(\nu)}a^2(ar)^{\nu-1}K_{\nu-1}(ar) & \mbox{if } r>0\\
-\frac{a^2}{2(\nu-1)} & \mbox{if } r=0,
 \end{cases}
\end{equation}
and
\begin{equation}\label{fun_G}
G_{\textrm{Mat}}(r; \nu, a)=\frac{2^{1-\nu}}{\Gamma(\nu)}a^4(ar)^{\nu-2}K_{\nu-2}(ar) \quad \mbox{if } r>0,
\end{equation}
where $\nu>1$.

\vspace{1cm}

\noindent\textbf{Appendix E: Proof of Proposition \ref{prop1}}

\vspace{0.5cm}

Let $\mathbf{Y}(\mathbf{s})=(Y_1(\mathbf{s}), Y_2(\mathbf{s}), Y_3(\mathbf{s}))^{\rm T}$ denote a tangential vector field 
in Cartesian coordinates with cross-covariance function $\mathbf{C}_{\mathrm{tan}, \mathbf{Z}}(\mathbf{s}, \mathbf{t})$. 
It can be converted to the canonical coordinates $(\hat{\mathbf{u}}, \hat{\mathbf{v}})$ by the transformation
$$
\mathbf{V}(\mathbf{s})\equiv(u(\mathbf{s}), v(\mathbf{s}))^{\rm T}=\mathbf{T}_{\mathbf{s}}\mathbf{Y}(\mathbf{s}),
$$
where
$$
\mathbf{T}_{\mathbf{s}}=\left( \begin{matrix}
-\sin\phi_{\mathbf{s}} & \cos\phi_{\mathbf{s}} & 0 \\
-\cos\theta_{\mathbf{s}}\cos\phi_{\mathbf{s}} & -\cos\theta_{\mathbf{s}}\sin\phi_{\mathbf{s}} & \sin\theta_{\mathbf{s}}
\end{matrix} \right).
$$
Then the cross-covariance function of $\mathbf{V}$ is
$$
\mathbf{C}_{\mathbf{V}}(\mathbf{s}, \mathbf{t})=\mathbf{T}_{\mathbf{s}}\mathbf{C}_{\mathrm{tan}, \mathbf{Z}}(\mathbf{s}, \mathbf{t})\mathbf{T}_{\mathbf{t}}^{\rm T}.
$$
Notice that
$$
\mathbf{T}_{\mathbf{s}} \mathbf{P}_{\mathbf{s}}
=\left( \begin{matrix}
-\sin\phi_{\mathbf{s}} & \cos\phi_{\mathbf{s}} & 0 \\
-\cos\theta_{\mathbf{s}}\cos\phi_{\mathbf{s}} & -\cos\theta_{\mathbf{s}}\sin\phi_{\mathbf{s}} & \sin\theta_{\mathbf{s}}
\end{matrix} \right),
$$
and
$$
\mathbf{T}_{\mathbf{s}}\mathbf{Q}_{\mathbf{s}}=
\left( \begin{matrix}
\cos\theta_{\mathbf{s}}\cos\phi_{\mathbf{s}} & \cos\theta_{\mathbf{s}}\sin\phi_{\mathbf{s}} & -\sin\theta_{\mathbf{s}}\\
-\sin\phi_{\mathbf{s}} & \cos\phi_{\mathbf{s}} & 0
\end{matrix} \right).
$$
Plugging in them to the expression of $\mathbf{C}_{\mathrm{tan}, \mathbf{Z}}(\mathbf{s}, \mathbf{t})$, we have
$$
\mathbf{C}_{\mathbf{V}}(\mathbf{s}, \mathbf{s})=-\left[\sigma_1^2F_{\textrm{Mat}}(0; \nu_1, a)
+\sigma_2^2F_{\textrm{Mat}}(0; \nu_2, a)\right]\mathbf{I}_2,
$$
which is constant with respect to $\mathbf{s}$.

\mjfan{By making use of (\ref{eqn_u}) and (\ref{eqn_v}), we can intuitively understand the conclusion that the co-located correlation between $u$ and $v$ has to be zero. Since $Z_1$ is isotropic, the change rates of the field at the same location along latitudinal and longitudinal directions are uncorrelated, i.e., $\mbox{Cov}(\partial Z_1/\partial \phi, \partial Z_1/\partial \theta)=0$ \citep[page 116, (5.7.3)]{Adler-09}. Similarly, the same holds for $Z_2$. Moreover, the cross-covariance between $Z_1$ and $Z_2$ is also isotropic. Due to the curvature of the sphere, for $Z_1$ and $Z_2$, the covariance of the change rates along latitudinal direction equals  the covariance of the change rates along longitudinal direction multiplied by $\sin ^2 \theta$, i.e.,
$\mbox{Cov}(\partial Z_1/\partial \phi, \partial Z_2/\partial \phi)=\mbox{Cov}(\partial Z_2/\partial \theta, \partial Z_1/\partial \theta)\sin^2 \theta$.}

\vspace{1cm}

\noindent\textbf{Appendix F: Proof of Proposition \ref{prop2}}

\vspace{0.5cm}

According to the construction of the TMM, for any $(\theta, \phi)$ except at the two poles,
\begin{eqnarray*}
u(\theta, \phi)\hat{\mathbf{u}}+v(\theta, \phi)\hat{\mathbf{v}}
&=& 
\nabla^{*}_{\mathbf{s}} \widetilde{Z}_1(\mathbf{s})+L^{*}_{\mathbf{s}} \widetilde{Z}_2(\mathbf{s}) \qquad \mathbb{P-}\mbox{a.e.}\\
&=& 
\left(\frac{1}{\sin{\theta}}\frac{\partial{\widetilde{Z}_1}}{\partial \phi}
+ \frac{\partial{\widetilde{Z}_2}}{\partial \theta} \right)\hat{\bm{\phi}}-\left(\frac{1}{\sin{\theta}}\frac{\partial{\widetilde{Z}_2}}{\partial \phi}-\frac{\partial{\widetilde{Z}_1}}{\partial \theta} \right)\hat{\bm{\theta}}\\
&=& 
\left(\frac{1}{\sin{\theta}}\frac{\partial{\widetilde{Z}_1}}{\partial \phi}+\frac{\partial{\widetilde{Z}_2}}{\partial \theta} \right)\hat{\mathbf{u}}
+\left(\frac{1}{\sin{\theta}}\frac{\partial{\widetilde{Z}_2}}{\partial \phi}-\frac{\partial{\widetilde{Z}_1}}{\partial \theta} \right)\hat{\mathbf{v}}.
\end{eqnarray*}
Thus, 
$$
u(\theta, \phi)=\frac{1}{\sin{\theta}}\frac{\partial{\widetilde{Z}_1}}{\partial \phi}
+ \frac{\partial{\widetilde{Z}_2}}{\partial \theta} \qquad \mathbb{P-}\mbox{a.e.},
$$
and 
$$
v(\theta, \phi)=\frac{1}{\sin{\theta}}\frac{\partial{\widetilde{Z}_2}}{\partial \phi}
-\frac{\partial{\widetilde{Z}_1}}{\partial \theta} \qquad \mathbb{P-}\mbox{a.e.}
$$
By condition \textbf{A1} and the chain rule for partial derivatives in quadratic mean 
\citep[Theorem 2.13]{Potthoff-10}, (\ref{eqn_u}) and (\ref{eqn_v}) follow.

We have assumed that the cross-covariance function of $(Z_1(\mathbf{s}), Z_2(\mathbf{s}))^{\rm T}$ (restricted on 
$\mathbb{S}^2$) only depends on the chordal distance, which is a function of 
$\theta_1, \theta_2$ and $\phi_1-\phi_2$. Moreover, the coefficients of the partial 
derivatives are independent of longitude. Using the same argument as \citet{JunS2008}, 
we conclude that $u$ and $v$ are axially symmetric both marginally and jointly.

\bibliography{refs}
\bibliographystyle{apalike}

\pagebreak
\begin{center}
\textbf{\large Supplementary Materials for ``Modeling Tangential Vector Fields on a Sphere"}
\end{center}
\setcounter{equation}{0}
\setcounter{figure}{0}
\setcounter{table}{0}
\setcounter{page}{1}
\makeatletter
\renewcommand{\theequation}{S\arabic{equation}}
\renewcommand{\thefigure}{S\arabic{figure}}

\vspace{1cm}

\textbf{S1: Implementation of the DFT for the cross-covariance matrix}

\vspace{0.5cm}

Suppose the observations are on a regular grid on the sphere $\{ (\theta_i, \phi_j): i=1,\cdots,n_{\rm lat}, j=1,\cdots,n_{\rm lon}\}$, where $\theta$ and $\phi$ represent the co-latitude and the longitude, respectively. Besides, $\theta_i-\theta_{i-1}=(\theta_{n_{\rm lat}}-\theta_1)/(n_{\rm lat}-1)$
and $\phi_j-\phi_{j-1}=2\pi/n_{\rm lon}$, $\phi_1=0$, $\phi_{n_{\rm lon}}=2\pi-2\pi/n_{\rm lon}$. Denote the observations as $(Y_1(\theta_i, \phi_j), Y_2(\theta_i, \phi_j))^{\rm T}, i=1,\cdots, n_{\rm lat}, j=1,\cdots, n_{\rm lon}$. We rearrange and stack them as $\left( \mathbf{Y}(\theta_1)^{\rm T}, \cdots, \mathbf{Y}(\theta_{n_{\rm lat}})^{\rm T}\right)^{\rm T}$, where
$$\mathbf{Y}(\theta_i)=\left(\mathbf{Y}_1(\theta_i)^{\rm T}, \mathbf{Y}_2(\theta_i)^{\rm T}\right)^{\rm T}=\left(
Y_1(\theta_i, \phi_1), \cdots, Y_1(\theta_i, \phi_{n_{\rm lon}}), Y_2(\theta_i, \phi_1), \cdots, Y_2(\theta_i, \phi_{n_{\rm lon}})
\right)^{\rm T}.$$
Since the cross-covariance function of the random field $\mathbf{Y}$ is axially symmetric,
\begin{dmath*}
\mbox{Cov}\left(\mathbf{Y}(\theta_i), \mathbf{Y}(\theta_j)\right)=\mathbb{E}\left( \mathbf{Y}(\theta_i) \mathbf{Y}(\theta_j)^{\rm T} \right)\\
=\left( \begin{matrix}
\mathbb{E}\left(\mathbf{Y}_1(\theta_i)\mathbf{Y}_1(\theta_j)^{\rm T}\right) & \mathbb{E}\left(\mathbf{Y}_1(\theta_i)\mathbf{Y}_2(\theta_j)^{\rm T}\right) \\
\mathbb{E}\left(\mathbf{Y}_2(\theta_i)\mathbf{Y}_1(\theta_j)^{\rm T}\right) & \mathbb{E}\left(\mathbf{Y}_2(\theta_i)\mathbf{Y}_2(\theta_j)^{\rm T}\right) \\
\end{matrix} \right)\\
=\left( \begin{matrix}
\left[C_{11}(\theta_i, \theta_j, \phi_k-\phi_l)\right]_{1 \leq k,l \leq n_{\rm lon}} & \left[C_{12}(\theta_i, \theta_j, \phi_k-\phi_l)\right]_{1\leq k,l \leq n_{\rm lon}}\\
\left[C_{21}(\theta_i, \theta_j, \phi_k-\phi_l)\right]_{1 \leq k,l \leq n_{\rm lon}} & \left[C_{22}(\theta_i, \theta_j, \phi_k-\phi_l)\right]_{1 \leq k,l \leq n_{\rm lon}}
\end{matrix}
\right)
=\left( \begin{matrix}
\mathbf{C}_{11}(\theta_i, \theta_j) & \mathbf{C}_{12}(\theta_i, \theta_j)\\
\mathbf{C}_{21}(\theta_i, \theta_j) & \mathbf{C}_{22}(\theta_i, \theta_j)
\end{matrix}\right),
\end{dmath*}
where $\mathbf{C}_{\cdot \cdot}(\theta_i, \theta_j)$ are circulant matrices, which have the form (omitting the subscripts $\cdot \cdot$)
$$\left( \begin{matrix}
c_0 & c_{n_{\rm lon}-1} & \cdots & c_1\\
c_1 & c_0 & \cdots & c_2\\
\vdots & \vdots & & \vdots\\
c_{n_{\rm lon}-1} & c_{n_{\rm lon}-2} & \cdots & c_0
\end{matrix} \right).$$
A circulant matrix can be diagonalized by the DFT matrix
$$\mathbf{F}\mathbf{C}_{\cdot \cdot}(\theta_i, \theta_j) \mathbf{F}^{-1}=\bm{\Lambda}_{\cdot\cdot},$$
where 
$$\mathbf{F}=\left( 
\begin{matrix}
1 & 1 & \cdots & 1\\
1 & \omega^1 & \cdots & \omega^{n_{\rm lon}-1}\\
1 & \omega^2 & \cdots & \omega^{2(n_{\rm lon}-1)}\\
\vdots & \vdots & & \vdots\\
1 & \omega^{n_{\rm lon}-1} & \cdots & \omega^{(n_{\rm lon}-1)(n_{\rm lon}-1)}
\end{matrix} \right),$$
and $\omega=\exp{(-2\pi i/n_{\rm lon})}$.
The diagonal matrix $\bm{\Lambda}_{\cdot \cdot}$ can be computed by
$$\bm{\Lambda}_{\cdot \cdot}=\mbox{diag}(\mathbf{F}\mathbf{c}_{\cdot\cdot}),$$
where $\mathbf{c}_{\cdot \cdot}=(c_0,\cdots, c_{n_{\rm lon}-1})^{\rm T}$.
The observations are transformed accordingly
$$\mathbf{Y}(\theta_i)=\left(\mathbf{Y}_1(\theta_i)^{\rm T}, \mathbf{Y}_2(\theta_i)^{\rm T}\right)^{\rm T} \longrightarrow \left(\widetilde{\mathbf{Y}}_1(\theta_i)^{\rm T}, \widetilde{\mathbf{Y}}_2(\theta_i)^{\rm T}\right)^{\rm T}=\left(\left(\mathbf{F}\mathbf{Y}_1(\theta_i)\right)^{\rm T}/\sqrt{n_{\rm lon}}, \left(\mathbf{F}\mathbf{Y}_2(\theta_i)\right)^{\rm T}/\sqrt{n_{\rm lon}}\right)^{\rm T},$$
such that 
\begin{dmath*}
\mathbb{E}\left(\widetilde{\mathbf{Y}}_{\cdot}(\theta_i)\widetilde{\mathbf{Y}}_{\cdot}(\theta_j)^{\rm H}\right)=\mathbf{F}\mathbb{E}\left( \mathbf{Y}_{\cdot}(\theta_i) \mathbf{Y}_{\cdot}(\theta_j)^{\rm T}\right)\mathbf{F}^{\rm H}/n_{\rm lon}\\
=\mathbf{F}\mathbb{E}\left( \mathbf{Y}_{\cdot}(\theta_i) \mathbf{Y}_{\cdot}(\theta_j)^{\rm T}\right)\mathbf{F}^{-1}
=\bm{\Lambda}_{\cdot \cdot},
\end{dmath*}
where $\cdot^{\rm H}$ represents the Hermitian transpose.
We rearrange the order of the observations
\begin{dmath*}
\left(\widetilde{Y}_1(\theta_1, \phi_1),\cdots,\widetilde{Y}_1(\theta_{n_{\rm lat}}, \phi_1),\cdots,\widetilde{Y}_1(\theta_1, \phi_{n_{\rm lon}}),\cdots, \widetilde{Y}_1(\theta_{n_{\rm lat}}, \phi_{n_{\rm lon}}),\\
\widetilde{Y}_2(\theta_1, \phi_1),\cdots,\widetilde{Y}_2(\theta_{n_{\rm lat}}, \phi_1),\cdots,\widetilde{Y}_2(\theta_1, \phi_{n_{\rm lon}}),\cdots, \widetilde{Y}_2(\theta_{n_{\rm lat}}, \phi_{n_{\rm lon}})\right)^{\rm T}.
\end{dmath*}
Then the corresponding cross-covariance matrix is
$$\bm{\Sigma}=\left( 
\begin{matrix}
\bm{\Sigma}_1 & \bm{\Sigma}_{12}\\
\bm{\Sigma}_{12}^{\rm H} & \bm{\Sigma}_2
\end{matrix}
\right),$$
where $\bm{\Sigma}_1, \bm{\Sigma}_{12}$ and $\bm{\Sigma}_2$ are complex block diagonal matrices.
The determinant of $\bm{\Sigma}$ can be computed by
$$\mbox{det}(\bm{\Sigma})=\mbox{det}(\bm{\Sigma}_1-\bm{\Sigma}_{12}\bm{\Sigma}_2^{-1}\bm{\Sigma}_{12}^{\rm H})\mbox{det}(\bm{\Sigma}_2),$$
and the inverse of $\bm{\Sigma}$ can be computed by 
$$\bm{\Sigma}^{-1}=\left(\begin{matrix}
(\bm{\Sigma}_1-\bm{\Sigma}_{12}\bm{\Sigma}_2^{-1}\bm{\Sigma}_{12}^{\rm H})^{-1} & -\bm{\Sigma}_1^{-1}\bm{\Sigma}_{12}(\bm{\Sigma}_2-\bm{\Sigma}_{12}^{\rm H}\bm{\Sigma}_1^{-1}\bm{\Sigma}_{12})^{-1}\\
(-\bm{\Sigma}_1^{-1}\bm{\Sigma}_{12}(\bm{\Sigma}_2-\bm{\Sigma}_{12}^{\rm H}\bm{\Sigma}_1^{-1}\bm{\Sigma}_{12})^{-1})^{\rm H} & (\bm{\Sigma}_2-\bm{\Sigma}_{12}^{\rm H}\bm{\Sigma}_1^{-1}\bm{\Sigma}_{12})^{-1}
\end{matrix}
\right),$$
where the inverses of $\bm{\Sigma}_1$ and $\bm{\Sigma}_2$ can be computed efficiently since they are both block diagonal matrices.

\newpage

\textbf{S2: Tangential vector fields derived from a full bivariate Mat\'ern model}

\vspace{0.5cm}

The underlying random field $\textbf{Z}$ is assumed to follow a full bivariate Mat\'ern model \citep{Gneiting-10}
\begin{equation*}\label{pars_m1}
C_{ii}(\lVert \mathbf{h} \rVert)=\sigma_i^2M(\lVert \mathbf{h} \rVert;\nu_i, a_i) \qquad \mbox{for } i=1,2,
\end{equation*}
and
\begin{equation*}\label{pars_m2}
C_{12}(\lVert \mathbf{h} \rVert)=C_{21}(\lVert \mathbf{h} \rVert)=\rho_{12} \sigma_1 \sigma_2 M(\lVert \mathbf{h} \rVert ;\nu_{12}, a_{12}).
\end{equation*}
It is valid if and only if 
\begin{equation}\label{full_mat_cond}
\rho_{12}^2 \leq \frac{\Gamma\left(\nu_1+\frac{d}{2}\right)}{\Gamma(\nu_1)}\frac{\Gamma\left(\nu_2+\frac{d}{2}\right)}{\Gamma(\nu_2)}
\frac{\Gamma(\nu_{12})^2}{\Gamma\left(\nu_{12}+\frac{d}{2}\right)^2}\frac{a_1^{2\nu_1}a_2^{2\nu_2}}{a_{12}^{4\nu_{12}}}\inf \limits_{t\geq 0}
\frac{(a_{12}^2+t^2)^{2\nu_{12}+d}}{( a_1^2+t^2 )^{\nu_1+(d/2)}(a_2^2+t^2)^{\nu_2+(d/2)}}.
\end{equation}
Condition (\ref{full_mat_cond}) implies that if $\nu_{12} < \frac{1}{2}(\nu_1+\nu_2)$, $\rho_{12}$ has to be zero, i.e., the two Cartesian components
of $\textbf{Z}$ are independent. Thus, we are mainly interested in the case when $\nu_{12} \geq \frac{1}{2}(\nu_1+\nu_2)$.

The derived cross-covariance function of $\mathbf{Y}_{\mathrm{tan}, \mathbf{Z}}(\mathbf{s})$ is
\begin{eqnarray}
&& \mathbf{C}_{\mathrm{tan}, \mathbf{Z}}(\mathbf{s}, \mathbf{t}) \nonumber\\
&=& -\left( \begin{matrix}
\sigma_1 \mathbf{P}_{\mathbf{s}} & \sigma_2 \mathbf{Q}_{\mathbf{s}}
\end{matrix} \right)\left( \begin{matrix}
\mathbf{K}(\mathbf{h}; \nu_1, a_1) & \rho_{12}\mathbf{K}(\mathbf{h};\nu_{12}, a_{12})\\
\rho_{12}\mathbf{K}(\mathbf{h};\nu_{12}, a_{12}) & \mathbf{K}(\mathbf{h}; \nu_2, a_2)
\end{matrix} \right)
\left( \begin{matrix}
\sigma_1 \mathbf{P}_{\mathbf{t}}^{\rm T} \\
 \sigma_2 \mathbf{Q}_{\mathbf{t}}^{\rm T}
\end{matrix} \right), \nonumber
\end{eqnarray}
where $\nu_1, \nu_2>1$.

\newpage
\textbf{S3: Accuracy of parameter estimation on irregular grids}

\vspace{0.5cm}

To investigate the accuracy of parameter estimation on irregular grids, we conduct some smaller scale experiments (due to the computational cost). Specifically, we generate $100$ realizations with 
$$
\bm{\theta}=(1, 1, 0.5, 3, 4, 1/2, 0.1, 0.1) 
$$
on two HEALPix grids \citep{Gorski-05} with the number of grid points being $192$ and $768$, respectively. The HEALPix grid partitions the unit sphere into equal area pixels, and thus is irregular.
\begin{figure}[htbp] 
   \centering
   \includegraphics[width=6in]{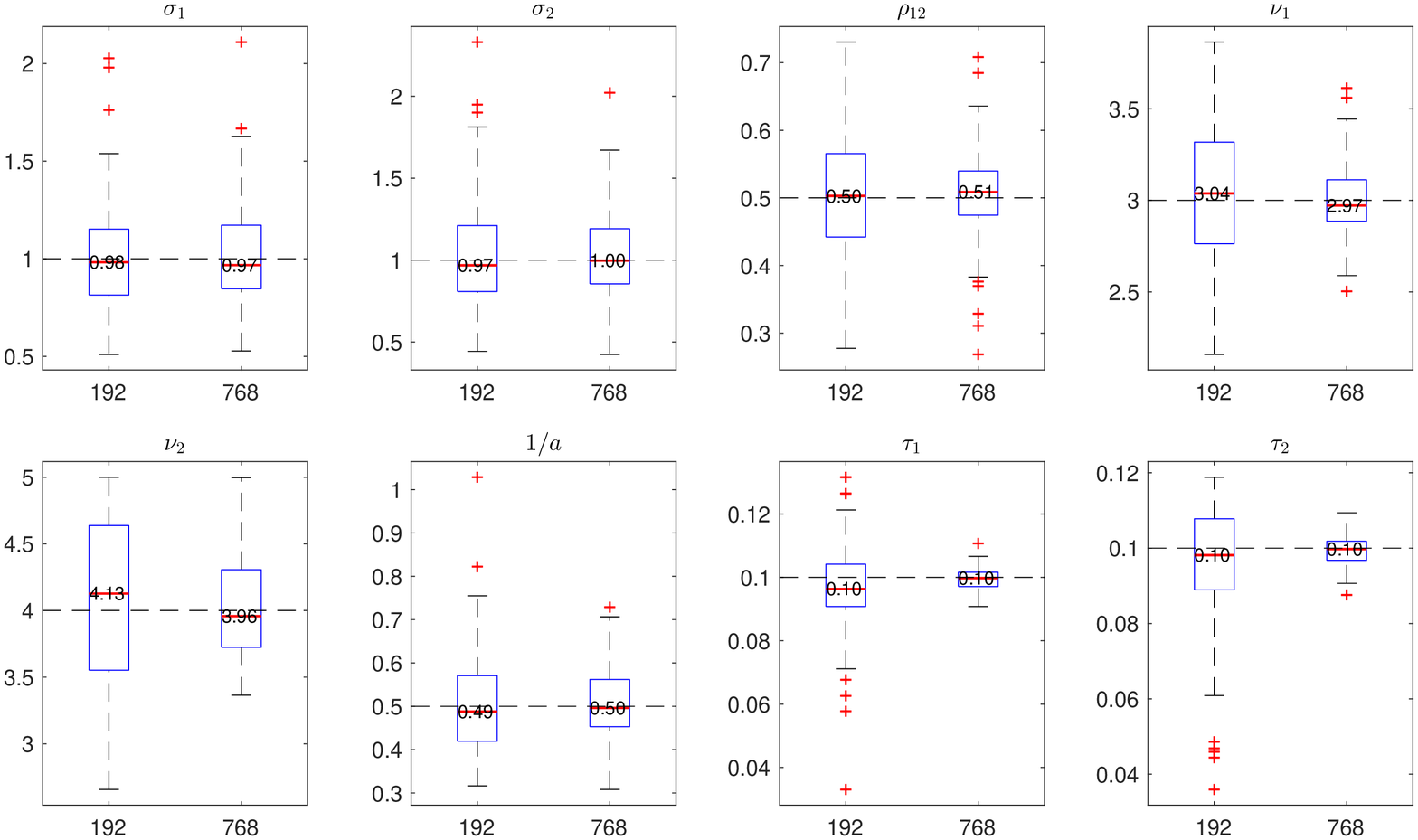}
   \caption{Results of the Monte Carlo simulation study to investigate the accuracy of 
   parameter estimation for the TMM \mjfan{on irregular grids.} The MLEs of 
   $\bm{\theta}=(\sigma_1, \sigma_2, \rho_{12}, \nu_1, \nu_2, 1/a, \tau_1, \tau_2)$
   are summarized by boxplots \mjfan{for two simulated data sets with increasing sample size (shown on x-axis)}.
   On each box, the central mark is the median (its value is explicitly 
   shown with two decimals), the edges of the box are the 25th and 75th percentiles, 
   the whiskers extend to the most extreme data points not considered outliers, and 
   outliers are plotted individually. The dashed horizontal lines are at the true values.}
   \label{fig:sim_MLE}
\end{figure}

\newpage
\textbf{S4: Accuracy of parameter estimation when the noise level varies}

\vspace{0.5cm}

\begin{figure}[htbp] 
   \centering
   \includegraphics[width=6in]{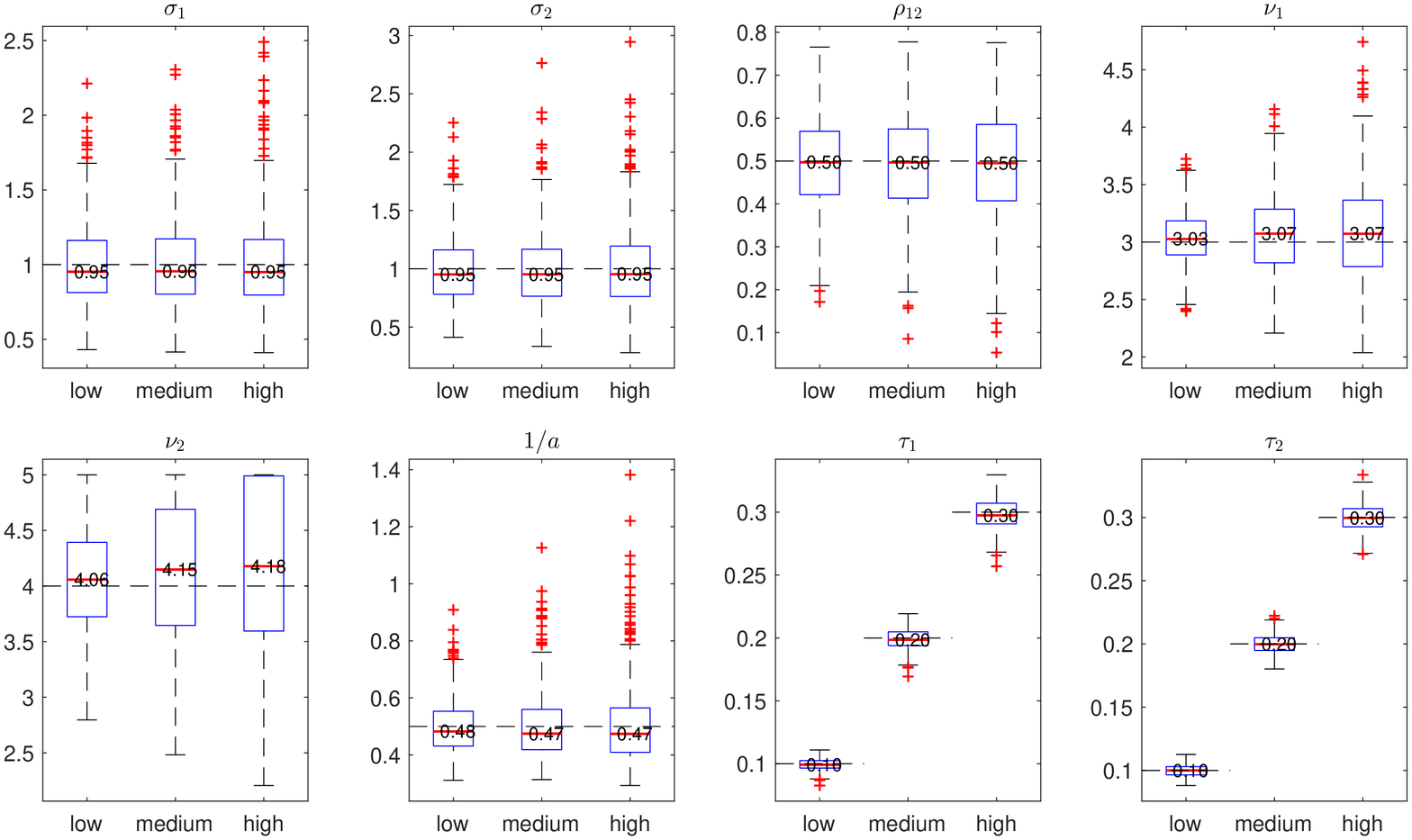}
   \caption{Results of the Monte Carlo simulation study to investigate the accuracy of 
   parameter estimation for the TMM \mjfan{when the noise level (shown on x-axis) varies.} The MLEs of 
   $\bm{\theta}=(\sigma_1, \sigma_2, \rho_{12}, \nu_1, \nu_2, 1/a, \tau_1, \tau_2)$
   are summarized by boxplots. 
   On each box, the central mark is the median (its value is explicitly 
   shown with two decimals), the edges of the box are the 25th and 75th percentiles, 
   the whiskers extend to the most extreme data points not considered outliers, and 
   outliers are plotted individually. The dashed horizontal lines are at the true values.}
   \label{fig:sim_MLE_noise}
\end{figure}

It is also interesting to see how the noise level affects the accuracy of parameter estimation. Specifically, we consider three different
cases of the noise level, and generate $500$ realizations with 
\begin{eqnarray}
\bm{\theta}=(1, 1, 0.5, 3, 4, 1/2, 0.1, 0.1) & \mbox{low noise} \nonumber\\
\bm{\theta}=(1, 1, 0.5, 3, 4, 1/2, 0.2, 0.2) & \mbox{medium noise} \nonumber\\
\bm{\theta}=(1, 1, 0.5, 3, 4, 1/2, 0.3, 0.3) & \mbox{high noise} \nonumber
\end{eqnarray}
on a regular grid with $(n_{\rm lat}, n_{\rm lon})=(15, 30)$, respectively. Figure \ref{fig:sim_MLE_noise} shows that the higher the noise level, the larger the standard errors of the estimates, especially
for $\nu_1$ and $\nu_2$.

\newpage

\textbf{S5: Accuracy of parameter estimation when the smoothness of the field varies}

\vspace{0.5cm}

\begin{figure}[htbp] 
   \centering
   \includegraphics[width=6in]{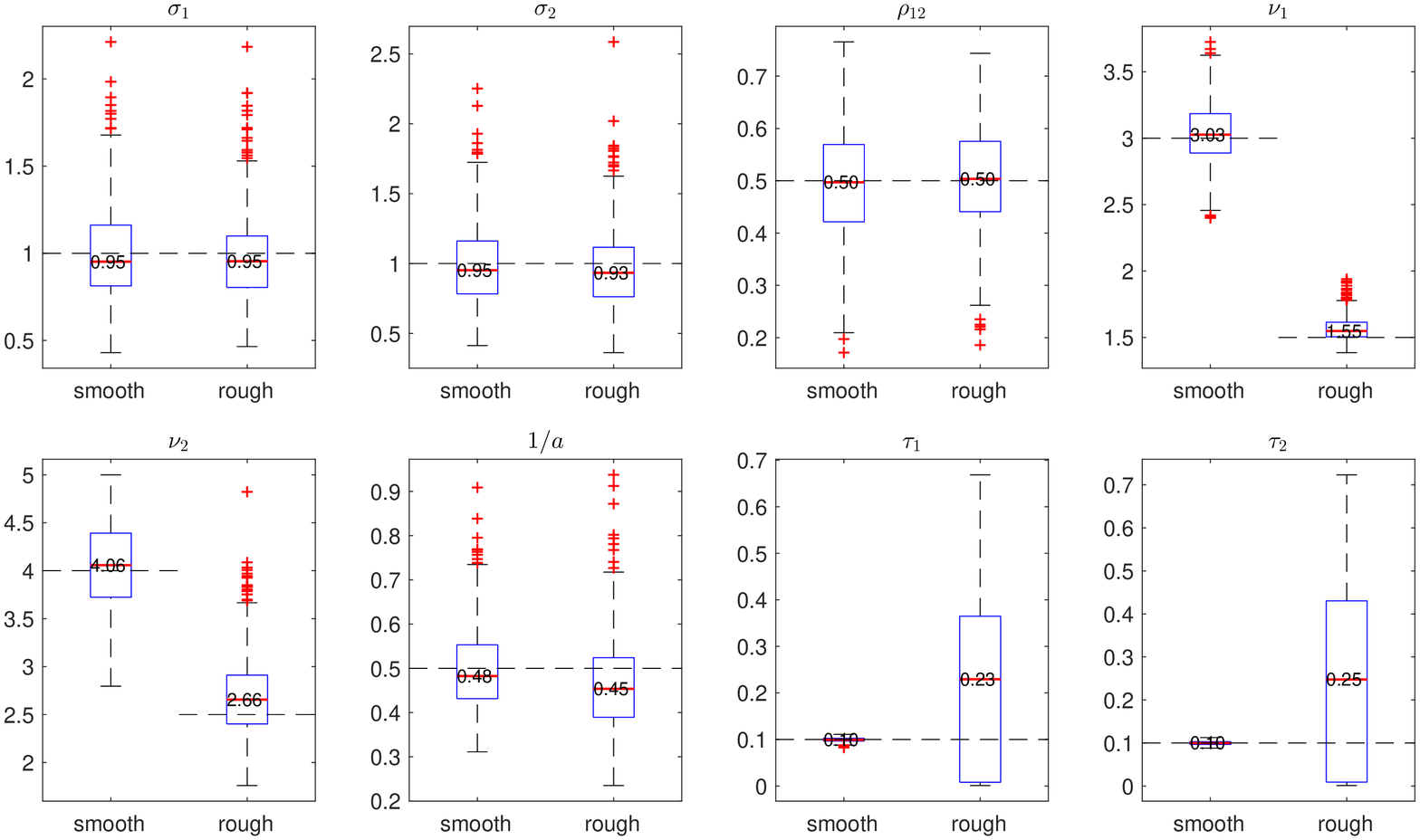}
   \caption{Results of the Monte Carlo simulation study to investigate the accuracy of 
   parameter estimation for the TMM \mjfan{when the smoothness of the field (shown on x-axis) varies.} The MLEs of 
   $\bm{\theta}=(\sigma_1, \sigma_2, \rho_{12}, \nu_1, \nu_2, 1/a, \tau_1, \tau_2)$
   are summarized by boxplots. 
   On each box, the central mark is the median (its value is explicitly 
   shown with two decimals), the edges of the box are the 25th and 75th percentiles, 
   the whiskers extend to the most extreme data points not considered outliers, and 
   outliers are plotted individually. The dashed horizontal lines are at the true values.}
   \label{fig:sim_MLE_rough}
\end{figure}

We also vary the smoothness of the field to test its effect on the accuracy of parameter estimation. Specifically, 
we generate $500$ realizations with 
\begin{eqnarray}
\bm{\theta}=(1, 1, 0.5, 3, 4, 1/2, 0.1, 0.1) & \mbox{smooth} \nonumber\\
\bm{\theta}=(1, 1, 0.5, 1.5, 2.5, 1/2, 0.1, 0.1) & \mbox{rough}. \nonumber
\end{eqnarray}
on a regular grid with $(n_{\rm lat}, n_{\rm lon})=(15, 30)$, respectively. Due to the differential operators applied, the latter specification gives a field as rough as the one generated by a Mat\'ern model with $\nu=1.5-1=0.5$. Figure \ref{fig:sim_MLE_rough} shows that it becomes more difficult to distinguish between the field and the white noise when the field is very rough, and thus $\tau_1$ and $\tau_2$ are significantly overestimated. Moreover, in this case, the upward bias of the estimates for $\nu_1$ and $\nu_2$ becomes more pronounced.

\newpage 

\textbf{S6: Accuracy of parameter estimation when covariates are included in the model}

\begin{figure}[htbp] 
   \centering
   \includegraphics[width=6in]{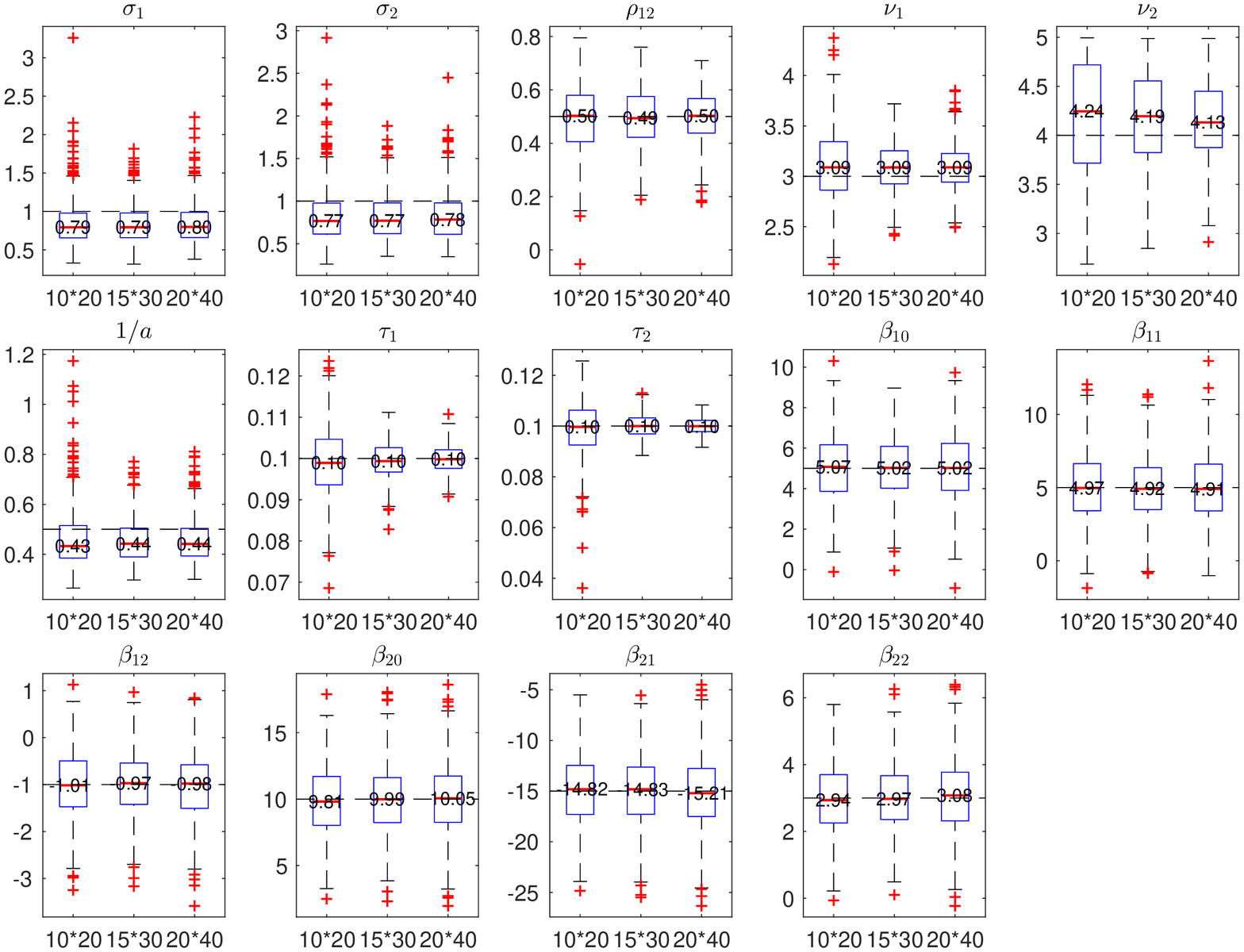}
   \caption{Results of the Monte Carlo simulation study to investigate the accuracy of 
   parameter estimation for the TMM \mjfan{when covariates are included in the model.} The MLEs of 
   $\bm{\theta}=(\sigma_1, \sigma_2, \rho_{12}, \nu_1, \nu_2, 1/a, \tau_1, \tau_2)$ and $\bm{\beta}_i, i=0,1,2$
   are summarized by boxplots \mjfan{for three simulated data sets with increasing sample size (shown on x-axis)}. 
   On each box, the central mark is the median (its value is explicitly 
   shown with two decimals), the edges of the box are the 25th and 75th percentiles, 
   the whiskers extend to the most extreme data points not considered outliers, and 
   outliers are plotted individually. The dashed horizontal lines are at the true values.}
   \label{fig:sim_MLE_covariate}
\end{figure}

We include the co-latitude as covariates in the model such that for the field $\textbf{Y}(\textbf{s}), \textbf{s}=(\theta, \phi)$, its mean is
$\bm{\beta}_0+\bm{\beta}_1\theta+\bm{\beta}_2\theta^2$, where $\bm{\beta}_i=(\beta_{1i}, \beta_{2i})^{\rm T}, i=0,1,2$ are two dimensional coefficient vectors, and its cross-covariance function still follows the TMM (augmented with nugget effects).
We generate $500$ realizations with 
$$\bm{\theta}=(1, 1, 0.5, 3, 4, 1/2, 0.1, 0.1),$$
and 
$$\bm{\beta}_0=(5, 10)^{\rm T}, \bm{\beta}_1=(5, -15)^{\rm T}, \bm{\beta}_2=(-1, 3)^{\rm T}$$
on three regular grids with $(n_{\rm lat}, n_{\rm lon}) = (10, 20), (15, 30)$ and $(20, 40)$, respectively. Figure \ref{fig:sim_MLE_covariate} shows that the estimates of $\sigma_1$, $\sigma_2$ and $1/a$ have larger bias than those obtained when no covariate is included. This suggests the difficulty in jointly estimating the parameters contained in both the mean and the random components. The biases of the estimates for $\bm{\beta}_i, i=0,1,2$ are small, while the standard errors of the estimates do not decrease significantly when the sample size increases.

\vspace{1cm}

\textbf{S7: Comparison between the empirical and fitted variances of the $u$ and $v$ residual fields}

\vspace{0.5cm}

\begin{figure}[htbp] 
   \centering
   \includegraphics[width=5in]{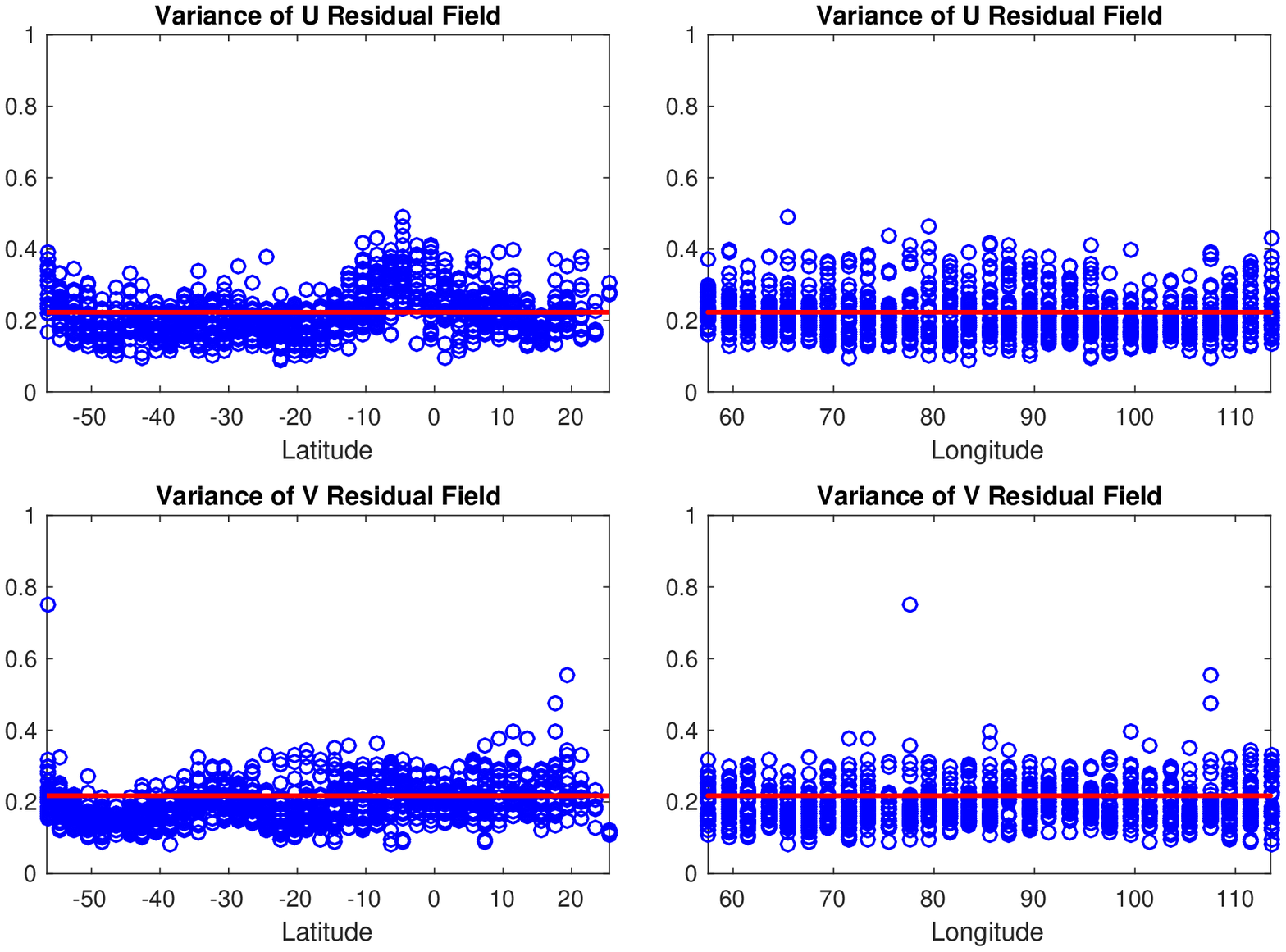}
   \caption{Empirical (circles) and fitted (solid line) variances of the $u$ and $v$ residual fields. They are plotted with respect to latitude (left) and longitude (right).}
   \label{fig:emp_fitted_var_uv}
\end{figure}

Due to Proposition 1, the fitted variances of the $u$ and $v$ residual fields are constant. Figure \ref{fig:emp_fitted_var_uv} shows that the fitted variances do not deviate from the empirical ones too much.

\end{document}